\documentclass[tradiabstract]{aa}

\usepackage{txfonts}
\usepackage{graphicx}
\usepackage{float}
\usepackage{natbib}
\usepackage[english]{babel}
\usepackage{color}

\begin{document}

\title{Dense-gas tracers and carbon isotopes in five 2.5$<$z$<$4 lensed dusty star-forming galaxies from the SPT SMG sample}

\author{M.~B\'ethermin\inst{1} \and T.~R.~Greve\inst{2} \and C.~De~Breuck\inst{3} \and J.~D.~Vieira\inst{4} \and M.~Aravena\inst{5} \and  S.~C.~Chapman\inst{6} \and Chian-Chou~Chen\inst{3} \and C.~Dong\inst{7} \and C.~C.~Hayward\inst{8} \and Y.~Hezaveh\inst{9} \and D.~P.~Marrone\inst{10} \and D.~Narayanan\inst{7,11,12} \and K.~A.~Phadke\inst{4} \and C.~A.~Reuter\inst{4} \and J.~S.~Spilker\inst{13} \and A.~A.~Stark\inst{14} \and M.~L.~Strandet\inst{15} \and A.~Wei\ss\inst{15}}

\institute{
Aix Marseille Univ, CNRS, CNES, LAM, Marseille, France, \email{matthieu.bethermin@lam.fr} \and
Department of Physics and Astronomy, University College London, Gower Street, London WC1E 6BT, UK \and
European Southern Observatory, Karl Schwarzschild Stra\ss e 2, 85748 Garching, Germany \and
Department of Astronomy and Department of Physics, University of Illinois, 1002 West Green St., Urbana, IL 61801 \and
N\'ucleo de Astronom\'{\i}a, Facultad de Ingenier\'{\i}a y Ciencias, Universidad Diego Portales, Av. Ej\'ercito 441, Santiago, Chile \and
Dalhousie University, Halifax, Nova Scotia, Canada \and
Department of Astronomy, University of Florida, Gainesville, FL 32611 USA \and
Center for Computational Astrophysics, Flatiron Institute, 162 Fifth Avenue, New York, NY 10010, USA \and
Kavli Institute for Particle Astrophysics and Cosmology, Stanford University, Stanford, CA 94305, USA \and
Steward Observatory, University of Arizona, 933 North Cherry Avenue, Tucson, AZ 85721, USA \and
University of Florida Informatics Institute, 432 Newell Drive, CISE Bldg E251, Gainesville, FL 32611 \and
Cosmic Dawn Centre (DAWN), University of Copenhagen, Julian Maries vej 30, DK-2100, Copenhagen, Denmark \and
Department of Astronomy, University of Texas at Austin, 2515 Speedway Stop C1400, Austin, TX 78712, USA \and
Harvard-Smithsonian Center for Astrophysics, 60 Garden Street, Cambridge, MA 02138, USA \and
Max-Planck-Institut f\"{u}r Radioastronomie, Auf dem H\"{u}gel 69 D-53121 Bonn, Germany
}

\date{Received ??? / Accepted ???}

\abstract{The origin of the high star formation rates (SFR) observed in
high-redshift dusty star-forming galaxies is still unknown. Large fractions of dense molecular gas
might provide part of the explanation, but there are few observational constraints on the amount of
dense gas in high-redshift systems dominated by star formation. In this paper,
we present the results of our Atacama large millimeter array (ALMA) program targeting dense-gas tracers
(HCN(5-4), HCO$^+$(5-4), and HNC(5-4)) in five strongly lensed galaxies from the
South Pole Telescope (SPT) submillimeter galaxy sample. We detected two of these
lines (S/N$>$5) in SPT-125-47 at z=2.51 and tentatively detected all three (S/N$\sim$3) in SPT0551-50 at z=3.16.
Since a significant fraction of our target lines is not detected, we developed a statistical method to derive
unbiased mean properties of our sample taking into account both detections and non-detections.
On average, the HCN(5-4) and HCO$^{+}$(5-4) luminosities of our sources are a
factor of $\sim$1.7 fainter than expected, based on the local L'$_{\rm
HCN(5-4)}$-L$_{\rm IR}$ relation, but this offset corresponds to only
$\sim$2\,$\sigma$ if we consider sample variance. We find that both the
HCO$^{+}$/HCN and HNC/HCN flux ratios are compatible with unity. The first ratio
is expected for photo-dominated regions (PDRs) while the second is consistent with PDRs or X-ray dominated regions (XDRs) and/or mid-infrared (IR)
pumping of HNC. Our sources are at the high end of the local relation between
the star formation efficiency, determined using the L$_{\rm IR}$/[CI] and L$_{\rm
IR}$/CO ratios, and the dense-gas fraction, estimated using the HCN/[CI] and
HCN/CO ratios. Finally, in SPT0125-47, which has the highest signal-to-noise
ratio, we found that the velocity profiles of the lines tracing dense (HCN,
HCO$^{+}$) and lower-density (CO, [CI]) molecular gas are similar. In addition
to these lines, we obtained one robust and one tentative detection of
$^{13}$CO(4-3) and found an average I$_{\rm ^{12}CO(4-3)}$/I$_{\rm
^{13}CO(4-3)}$ flux ratio of 26.1$_{-3.5}^{+4.5}$, indicating a young but not pristine interstellar medium. We argue that the combination of large and slightly enriched gas reservoirs and high dense-gas fractions could explain the prodigious star formation in these systems.}


\titlerunning{Dense-gas tracers and carbon isotopes in five lensed DSFGs}

\authorrunning{B\'ethermin et al.}

\maketitle

\section{Introduction}
Traditionally, the molecular gas in high-redshift galaxies has been probed by
observations of the rotational lines of CO \citep[e.g.,][]{Solomon2005}.
Systematic CO line surveys have shown an increasing molecular gas fraction with
redshift
\citep[e.g.,][]{Tacconi2010,Tacconi2013,Saintonge2013,Carilli2013,Dessauges2015,Aravena2016,Keating2016}.
Other methods based on the galaxy dust content found similar results
\citep{Magdis2012b,Scoville2014,Bethermin2015a,Scoville2016,Schinnerer2016}.
These analyses suggest that high star formation activity in massive galaxies
observed at high redshift is related to their larger gas content. Furthermore, an increase in the star formation efficiency, that is, the
star formation rate relative to the total gas mass (as traced by CO or
dust), seems necessary to account for the prodigious star formation in the most
extreme systems
\citep[e.g.,][]{Engel2010,Genzel2010,Daddi2010b,Magdis2011,Tan2014,Hodge2015,Silverman2015}.

In contrast, we have much less information about the amount of dense, actively
star-forming gas in high-redshift galaxies. The dense gas is usually traced using
the rotational lines of high-dipole molecules such as HCN, which have H$_{2}$
critical densities $\gtrsim 10^4 {\rm cm}^{-3}$ \citep{Shirley2015}. While \citet{Kauffmann2017} argue that in some Galactic clouds the HCN(1-0) line traces gas densities similar to those traced by high-J CO lines ($\sim 10^{3}\,{\rm cm}^{-3}$),
the higher J transitions of HCN are still considered to be our best indicators of
dense gas in galaxies.


As gas in the interstellar medium collapses to form stars, it passes
though the density regime where the hydrogen is predominantly molecular and is
dense enough to excite rotational transitions of HCN and HCO$^+$ into emission
\citep{larson94}.  For stars to form, the density must be sufficiently high in
the star-forming cloud cores that self-gravity dominates over tidal shear.  For
much of the volume of a typical galaxy, the threshold density for collapse in
spite of tidal shear and the threshold density for excitation of high dipole
moment molecules into emission are approximately the same \citep{stark89}.  The
luminosity of the HCN line is then a good measure of the star formation rate.
In the central kiloparsec of large galaxies, however, the density of the
interstellar medium may be high enough to excite HCN but not high enough to
resist the very much higher tidal shear in those regions of high differential
rotation; the HCN line may then have a large beam-filling factor and be bright
in emission from an extended, non-cloudy molecular gas, even though no star
formation can take place.

In the local Universe, \citet{Gao2004} found that in log-log space, the HCN(1-0) luminosity is linearly
proportional to star formation rate (SFR), as gauged by the total infrared
luminosity (L$_{\rm IR}$), with a small scatter over $2.5$ decades in
luminosity. The interpretation of this result is that the star formation rate in
galaxies is primarily controlled by the dense gas fraction. Linear luminosity
relations have been found using CO(3-2), HCO$^+$(1-0), HCO$^+$(3-2),
HCO$^+$(5-4), HCN(4-3), CS(5-4), CS(7-6), and formaldehyde dense observations of
a subset of the same galaxies
\citep{yao03a,narayanan05a,graciacarpio06a,mangum08a,iono09a,Juneau2009,wang11a,mangum13a,Zhang2014}.

Observations of Milky Way clumps in HCN(1-0), as well as in a variety of
other dense gas tracers, by \citet{wu05a,wu10a} found a roughly linear relation
between the SFR and the dense-gas mass (M$_{\rm dense}$) consistent with the
galaxy-integrated measurements.  They arrive at the same interpretation as
\citet{Gao2004}, namely that the SFR of a galaxy scales linearly with the number
of dense-gas ``units'' in a galaxy, with the dense-gas star formation efficiency
being constant.  Similarly, observations toward high visual extinction lines of
sight by \citet{lada10a,lada12a} and \citet{heiderman10a} support this
interpretation. In contrast, \citet{Bigiel2016} showed that the increase in dense-gas fraction towards the center of M\,51 leads to a
decrease in the dense-gas star formation efficiency ($\propto$\,L$_{\rm
IR}$/L'$_{\rm HCN}$), since there will be fewer overdense regions able to
gravitationally collapse if the average density is higher.

This interpretation, however, as well as the claimed linearity of the 
SFR-dense gas relation is debated in the literature.
For example, \citet{bussmann08a} observe sub-linear IR-HCN(3-2)
luminosity relations for nearby galaxies, while 
\citet{Garcia-Burillo2012} found a slightly super-linear IR-HCN(1-0) luminosity relation.
Both of these sets of observations can be reconciled with simulations 
\citep[e.g.,][]{narayanan08b,narayanan11a}.
Similarly, star formation models suggest that the
SFR-dense gas relationship is primarily set by the gas density probability
distribution function (PDF) in galaxies.  As a result, a range of SFR-dense gas
mass slopes may be expected, depending on the effective density of the tracer
and the exact form of the gas density PDF
\citep[e.g.,][]{Krumholz2007,narayanan08b,hopkins13a,Popping2014b,onus18a}.
In a high-pressure turbulent
interstellar medium (ISM), the average gas density is expected to be higher at
all physical scales with a larger fraction of the total gas mass residing at
higher densities ($> 10^4\,{\rm cm^{-3}}$). This type of high-pressure turbulent
ISM is more likely to occur in vigorously star-forming regions than in more
quiescent ISM conditions \citep{Papadopoulos2010}. This is supported by many
numerical simulations, which tend to predict a dramatic increase in the
dense-gas fraction during major mergers causing a short ($\sim$100\,Myr) boost
of the star formation efficiency \citep[e.g.,][]{Renaud2014}. However, a recent
study by \citet{Fensch2017} suggests that this phenomenon could not be efficient
in mergers of two gas-rich galaxies at high redshift. Observational constraints
are thus essential to test this result. HCN, HCO$^{+}$, and CS studies of local (ultra-)luminous infrared galaxies (ULIRGS) have found high gas fractions \citep[e.g.,][]{Gao2004,Garcia-Burillo2012,Zhang2014} -- but so far this has not
been firmly established at high redshift.


The detections of dense-gas tracers such as HCN and HCO$^+$ remain scarce at
high redshift due to the faintness of these lines (typically more than
10$\times$ fainter than CO). In fact, all such detections to date were obtained
with the assistance of gravitational lensing. Early efforts by, for example,
\citet{Greve2006} and \citet{Gao2007} to detect HCN(1-0) in high-z starburst
galaxies resulted only in upper limits. Most of the firm detections were
obtained for quasars: for example, HCN(1-0) in the Cloverleaf at $z = 2.6$ by
\citealt{Solomon2003}, IRAS F10214+4724 at $z = 2.3$ by \citealt{VandenBout2004},
and J1409+5628 at $z = 2.6$ by \citealt{Carilli2005} and HCN(5-4) and HCN(6-5) in
APM08279+5255 at $z = 3.9$ \citep{Barvainis1997,Wagg2005,Weiss2007,Riechers2010}. There are even fewer
published detections of star-formation-dominated objects. The HCN(3-2) line was
detected in SMMJ1213511-0102 at $z=2.3$ by \citet{Danielson2011} and recently
\citet{Oteo2017} reported two new detections of HCN(3-2) and HCO$^{+}$(3-2) in
SDP.9 and SDP.11 at $z=1.6$ and $z=1.8$, respectively (they also detected the 1-0
transition of these two molecules in SDP.9). So far, no detection at z$>$2.5 has been
reported in systems dominated by star formation. However, \citet{Spilker2014}
stacked all the cycle-0 Atacama large millimeter array (ALMA) spectral scans of the South Pole telescope submillimeter galaxy sample (SPT SMG, z$\sim$3.9) and detected HCN(4-3), HCO$^{+}$(4-3), and HCO$^{+}$(6-5).

In this paper, we present deep ALMA observations of HCN(5-4) in a sample
of five lensed dusty star-forming galaxies (DSFGs) between $z=2.5$ and $z=4$
from the South Pole Telescope (SPT) sample \citep{Vieira2010,Vieira2013}. 
We chose to target HCN(5-4), since the HNC(4-3) line is not observable beyond $z=3.2$ and we
want to observe the same transition in all the sources of our sample. The
transition has a typical effective excitation density of $\sim$10$^6$\,cm$^{-3}$
\citep{Shirley2015} and is therefore a {\it bone fide} tracer of the densest
molecular gas in galaxies. In addition to HCN(5-4), and since they can be observed at the same time, our observations also
targeted the HCO$^{+}$(5-4) and HNC(5-4) lines, which are dense-gas tracers in their own right. However, these two lines are harder to interpret
because of their more complex chemistry. HCO$^{+}$, being an ion, is dependent
on the ionization of the dense molecular clouds and is thus a less direct tracer of
these high densities \citep{Papadopoulos2007}. The line ratio between HCN and
HNC is dependent on the physical conditions and varies from $\sim$1 in dense
dark clouds \citep{Hirota1998} to $\sim$10$^{-2}$ in hot and dense regions of
star formation like Orion \citep{Schilke1992}. In addition to the above lines, we
were able to simultaneously target the $J=4-3$ transition of the $^{13}$CO
isotopologue in four of our sources. $^{13}$C is a secondary nucleus, which
is not produced from fusion of hydrogen or helium in massive short-lived stars. Thus, detecting this line is
a clue of previous star formation episodes in these galaxies \citep{Hughes2008,Henkel2010}.

In Sect.\,\ref{sect:data} we present our observations and our data reduction and line extraction methods. We then discuss the properties of dense-gas tracers in Sect.\,\ref{sect:dense_gas_tracers}. In Sect.\,\ref{sect:comp_gas_tracers}, we compare the properties of dense and lower density molecular gas tracers (CO, [CI]). Finally, we present and interpret our measurements of the CO isotopes in Sect.\,\ref{sect:isotopes}.
In this paper, we assume a \citet{Planck2015_cosmo} cosmology and a \citet{Chabrier2003} initial mass function (IMF).

\section{Data}

\label{sect:data}

\begin{table*}
\centering
\caption{\label{tab:obs} Summary of the characteristics of our observations. Each source was observed in a separate ALMA scheduling block (SB). Two similar SBs were used for SPT0551-50, which required a longer integration. The total observing time and the time on source are t$_{\rm obs}$ and t$_{\rm on}$, respectively. PWV is the average precipitable water vapor level during our observations. The beam size in the table was derived using a natural weighting.}
\begin{tabular}{lccccccccc}
\hline
\hline
Source & RA & Dec & Observing date & t$_{\rm obs}$ & t$_{\rm on}$ & PWV & band & N$_{\rm antennae}$& Synthesized beam size\\
\hline
SPT0103-45 & 01:03:11 & -45:38:53 & 2017-01-08 & 41\,min &18\,min & 4.8\,mm & 3 & 44 & 3.11"$\times$2.92" \\
SPT0125-47 & 01:25:07 & -47:23:56 & 2017-01-09 & 26\,min & 13\,min & 4.2\,mm & 4 & 44 & 2.52"$\times$1.96" \\
SPT0125-50 & 01:25:48 & -50:38:21 & 2016-12-29 & 60\,min & 33\,min & 2.8\,mm & 3 & 48 & 2.66"$\times$2.37" \\
SPT0300-46 & 03:00:04 & -46:21:24 & 2017-01-08 & 75\,min & 45\,min & 5.2\,mm & 3 & 41 & 3.44"$\times$2.72" \\
SPT0551-50 & 05:51:39 & -50:58:01 & 2017-01-07 & 86\,min & 54\,min & 5.5\,mm & 3 & 44 & 3.22"$\times$2.35" \\
\hline
\end{tabular}
\end{table*}

\subsection{Observations}

\label{sect:obs}

This paper is based on ALMA cycle-4 observations (2016.1.00065.S, PI:
B\'ethermin) of five DSFGs from the SPT sample
\citep{Vieira2013,Weiss2013,Strandet2016}, which were selected for having bright
apparent infrared luminosities (L$_{\rm IR} \ge 5 \times 10^{13}\,{\rm
L_{\odot}}$) and high-quality ancillary data.  ALMA bands 3 and, when needed, 4
were tuned to the redshifted frequencies of the HCN(5-4) transition. When possible,
spare spectral windows of the correlator were placed such as to cover
$^{13}$CO(4-3), HCO$^{+}$(5-4), HNC(5-4), and [CI](1-0) (this latest line is only observable in
SPT0125-47). Each spectral window covers 1.875\,GHz and contains 240 channels
(coarsest frequency domain mode resolution with an online spectral averaging by a factor of 16).
The previously mentioned lines are covered by two contiguous spectral windows in
one of the side bands.

The sensitivities of the observations were determined to reach 5\,$\sigma$
(10\,$\sigma$ for SPT0125-47) using the L'$_{\rm HCN}$-L$_{\rm IR}$ of
\citet{Zhang2014}. Since our sources are gravitationally lensed and have an
extension of $\sim$1\,arcsec \citep{Spilker2016}, we requested the most compact
configuration of the array to avoid spreading the flux of our sources over
several synthesized beams and thus maximizing our chances to detect the
integrated emission of our objects. The characteristics of our survey are
summarized in Table\,\ref{tab:obs}. The observations were performed on
December 29th, 2016 for SPT0125-47 and between January 7th, 2017 and January 9th, 2017 for the other
sources. During our observations, the precipitable water vapor level (PWV) was
between 2.8 and 5.5\,mm.


\begin{figure*}
\centering
\begin{tabular}{cc}
\includegraphics[width=8.5cm]{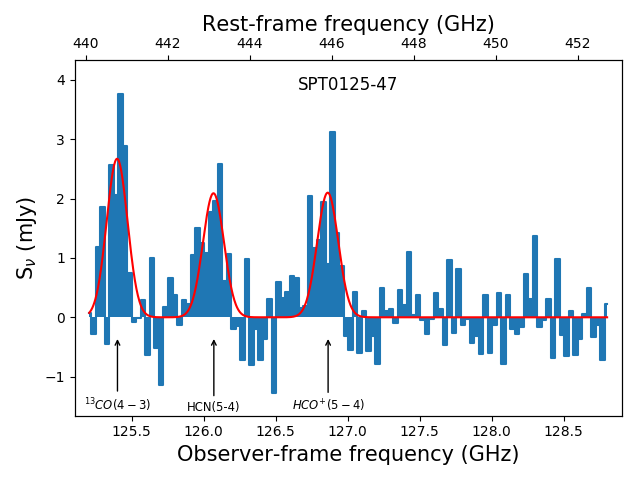} & \includegraphics[width=8.5cm]{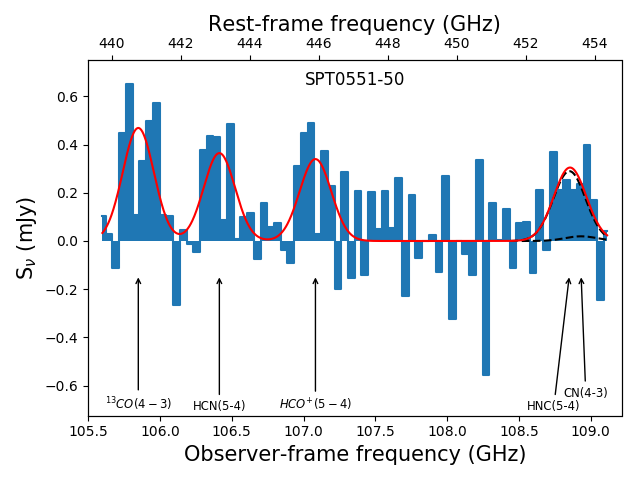}\\
\includegraphics[width=8.5cm]{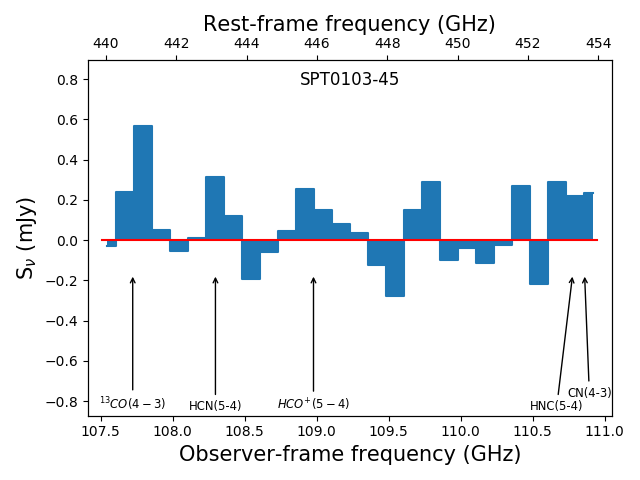} & \includegraphics[width=8.5cm]{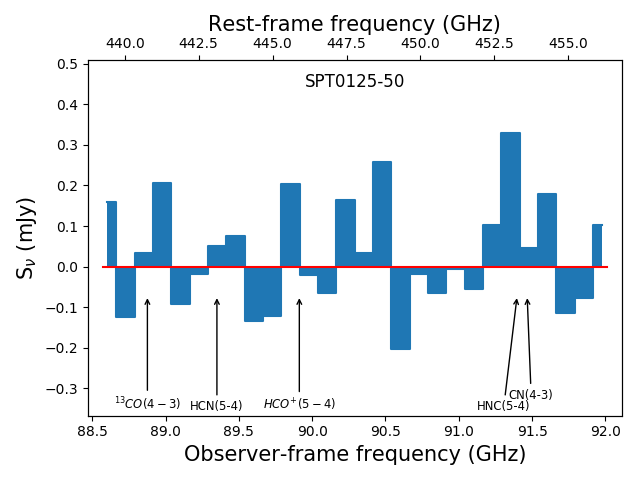}\\
\end{tabular}
\includegraphics[width=8.5cm]{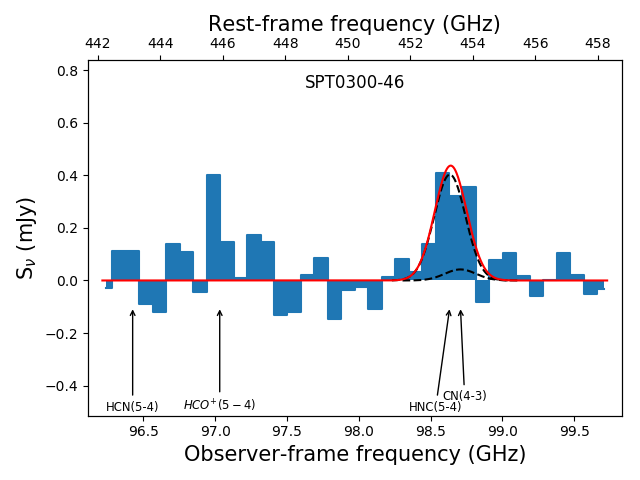}
\caption{\label{ref:spectra} Continuum-subtracted spectra of our five sources (blue filled histograms). Two spectral windows were combined to produce them. The best fit of the baseline is subtracted from the data. The red solid line is our best-fit model and the black dashed line is the best-fit decomposition of the blend between HNC(5-4) and CN(4-3). Only lines with S/N$>$3 are shown. The S/N of the blend between HNC(5-4) and CN(4-3) in SPT0551-50 only has  an S/N of 2.6 using our deblending method, which could be underestimated. However, its S/N determined in Appendix\,\ref{sect:fit_vs_mom0} using the moment-zero map is 3.3. We thus included the line in the plot. The three bottom panels show sources with a low S/N. Their spectra were rebinned by a factor of two in the figure to allow a better visualization of the faint lines.}
\end{figure*}

\subsection{Data reduction and extraction of the spectra}

\label{sect:data_red}

We analyzed our data with the CASA software \citep{McMullin2007}. They were initially calibrated by the standard ALMA pipeline at the observatory. In addition, we manually flagged some antennae and spectral windows with poor bandpass, phase, or amplitude calibration. In particular, the antenna DA48 had y and z position offsets larger than 2\,cm in all the January 2017 observations and its data were flagged systematically. The overall quality of the rest of the data is very good.

The data were imaged using the \textrm{CLEAN} algorithm, and a natural weighting
scheme was chosen to maximize the point-source sensitivity. We used cleaning
thresholds corresponding to 3\,$\sigma$ in a given channel using the map standard deviation around our source to estimate the noise. Since our lines are
very broad and the signal-to-noise ratio (S/N) is low, the channels were again rebinned at the imaging
stage. Because SPT0125-47 has several 5\,$\sigma$ detections, we only needed a
rebinning by a factor of four. Similarly, since SPT0300-46 and SPT0551-50 have at
least one line above 3\,$\sigma$, we used a factor of six. For the remaining
sources, we set the rebinning factor to eight.

We decided not to subtract the continuum directly in the uv plane. Instead, the continuum
was subtracted later at the line extraction stage (see
Sect.\,\ref{sect:line_ext}). This choice was motivated by the presence of
numerous broad lines in the two contiguous spectral windows of interest.  The
only good area of the spectrum without line contamination is between the HCO$^+$
and the HNC line, but is too narrow to accurately constrain the slope of the
continuum (see discussion in Appendix\,\ref{sect:fit_vs_mom0}). The imaging of the other side band, which is free of detected lines
except for SPT0125-47, showed that, at the depth of our observations, the
continuum varies significantly over 2$\times$1.875\,GHz. It is thus necessary to
use a first-order subtraction of the continuum.

Before extracting the spectra, we checked that our sources are not significantly extended. We produced high S/N images by combining all the channels of the four spectral windows (mode multi-frequency synthesis). We fitted a two-dimensional Gaussian model and found the width of the Gaussian to be consistent with that of the synthesized beam. Since the continuum emission of our sources is compatible with a point source at our ALMA resolution, we thus extracted their spectrum from the ALMA cube at the centroid of this Gaussian model. Using this method, we implicitly assume that the size of the regions emitting the dense-gas lines is not much more extended than the continuum. The extracted spectrum corresponds to the two contiguous spectral windows in the side band of HCN(5-4) and is presented in Fig.\,\ref{ref:spectra}.

\begin{table*}
\centering
\caption{\label{ref:lineext} Summary of the properties of our sources. We report the properties of lines of our detections and tentative detections (S/N$\ge$3). Upper limits correspond to 3 $\sigma$. The blend between HNC(5-4) and CN(4-3) in SPT0551-50 can be considered as a tentative detection as justified in Fig.\,\ref{ref:spectra} and Appendix\,\ref{sect:fit_vs_mom0}.}
\begin{tabular}{lrrrrr}
\hline
\hline
Source  & SPT0103-45\tablefootmark{a} & SPT0125-47 & SPT0125-50\tablefootmark{a} & SPT0300-46\tablefootmark{a} & SPT0551-50 \\ 
 \hline 
z & 3.0917 & 2.5148 & 3.9593 & 3.5955 & 3.1641 \\ 
Magnification $\mu$\tablefootmark{b} & 14.4 & 18.9 & 26.7 & 14.0 & 6.3 \\ 
Intrinsic 8-1000$\mu$m luminosity L$_{\rm IR}$ (10$^{12}$ L$_\odot$) & 6.19$\pm$1.67 & 9.63$\pm$0.90 & 3.92$\pm$0.36 & 5.67$\pm$1.24 & 10.02$\pm$1.52 \\ 
\hline 
\multicolumn{6}{c}{$^{13}$CO(4-3)} \\ 
\hline 
Observed frequency $\nu_{\rm obs}$ (GHz) & 107.722 & 125.402 & 88.876 & -   & 105.849 \\ 
Velocity offset $\Delta v$ (km/s) & -   & 6$\pm$19\tablefootmark{e} & -   & -   & -65$\pm$92 \\ 
Full width at half maximum (km/s) & -   & 407$\pm$54\tablefootmark{e} & -   & -   & 717$\pm$235 \\ 
Peak flux density (mJy) & <1.64 & 2.67$\pm$0.37\tablefootmark{e} & <0.47 & -   & 0.42$\pm$0.12 \\ 
S/N & 2.7 & 7.2 & 0.1 & -   & 3.4 \\ 
Line flux I$_{\rm ^{13}CO(4-3)}$ (Jy km/s) & <0.73 & 1.15$\pm$0.22 & <0.16 & -   & 0.32$\pm$0.15 \\ 
Luminosity L'$_{\rm ^{13}CO(4-3)}$ (10$^8$ K km/s pc$^2$) & <15.0 & 12.8$\pm$2.4 & <2.7 & -   & 15.7$\pm$7.1 \\ 
Average sample Luminosity L'$_{\rm ^{13}CO(4-3)}\tablefootmark{c}$ & \multicolumn{5}{c}{12.3$\pm$2.7} \\ 
\hline 
\multicolumn{6}{c}{HCN(5-4)} \\ 
\hline 
Observed frequency $\nu_{\rm obs}$ (GHz) & 108.296 & 126.071 & 89.351 & 96.425 & 106.413 \\ 
Velocity offset $\Delta v$ (km/s) & -   & 39$\pm$45 & -   & -   & 74$\pm$89 \\ 
Full width at half maximum (km/s) & -   & 475$\pm$128 & -   & -   & 529$\pm$232 \\ 
Peak flux density (mJy) & <1.28 & 1.90$\pm$0.36 & <0.95 & <0.39 & 0.39$\pm$0.13 \\ 
S/N & 1.8 & 5.3 & 1.8 & 0.5 & 3.0 \\ 
Line flux I$_{\rm HCN(5-4)}$ (Jy km/s) & <0.55 & 0.96$\pm$0.23 & <0.27 & <0.32 & 0.22$\pm$0.11 \\ 
Luminosity L'$_{\rm HCN(5-4)}$ (10$^8$ K km/s pc$^2$) & <11.2 & 10.6$\pm$2.5 & <4.5 & <8.6 & 10.6$\pm$5.1 \\ 
Average sample Luminosity L'$_{\rm HCN(5-4)}\tablefootmark{c}$ & \multicolumn{5}{c}{7.1$\pm$1.6} \\ 
\hline 
\multicolumn{6}{c}{HCO$^{+}$(5-4)} \\ 
\hline 
Observed frequency $\nu_{\rm obs}$ (GHz) & 108.977 & 126.864 & 89.912 & 97.031 & 107.083 \\ 
Velocity offset $\Delta v$ (km/s) & -   & 61$\pm$45 & -   & -   & 170$\pm$57 \\ 
Full width at half maximum (km/s) & -   & 496$\pm$119 & -   & -   & 318$\pm$194 \\ 
Peak flux density (mJy) & <0.92 & 1.92$\pm$0.35 & <0.87 & <0.63 & 0.50$\pm$0.16 \\ 
S/N & 1.1 & 5.6 & 1.5 & 1.7 & 3.1 \\ 
Line flux I$_{\rm HCO^{+}(5-4)}$ (Jy km/s) & <0.37 & 1.01$\pm$0.20 & <0.25 & <0.52 & 0.17$\pm$0.07 \\ 
Luminosity L'$_{\rm HCO^{+}(5-4)}$ (10$^8$ K km/s pc$^2$) & <7.6 & 11.1$\pm$2.2 & <4.0 & <13.8 & 8.1$\pm$3.4 \\ 
Average sample Luminosity L'$_{\rm HCO^{+}(5-4)}\tablefootmark{c}$ & \multicolumn{5}{c}{6.7$\pm$1.3} \\ 
\hline 
\multicolumn{6}{c}{HNC(5-4) and CN(4-3) (blended)} \\ 
\hline 
Obs. HNC(5-4) frequency $\nu_{\rm obs}$ (GHz) & 110.778 & -   & 91.398 & 98.634 & 108.852 \\ 
Obs. CN(4-3) frequency $\nu_{\rm obs}$ (GHz) & 110.860 & -   & 91.466 & 98.707 & 108.933 \\ 
Full width at half maximum (km/s) & -   & -   & -   & 780$\pm$212 & 726$\pm$165 \\ 
S/N of the blend\tablefootmark{d} & 2.3 & -   & 2.5 & 3.5 & 2.6 \\ 
Total blended flux I$_{\rm tot}$ (Jy km/s) & <0.58 & <0.00 & <0.45 & 0.37$\pm$0.11 & 0.25$\pm$0.09 \\ 
Blended lum. L'$_{\rm tot}$ (10$^8$ K km/s pc$^2$) & <11.4 & <0.0 & <7.0 & 9.6$\pm$2.8 & 11.5$\pm$4.4 \\ 
Average sample Luminosity L'$_{\rm HNC(5-4)}\tablefootmark{c}$ & \multicolumn{5}{c}{4.8$\pm$1.6} \\ 
Average sample Luminosity L'$_{\rm CN(4-3)}\tablefootmark{c}$ & \multicolumn{5}{c}{2.2$\pm$1.4} \\ 
\hline
\end{tabular}
\tablefoot{\tablefoottext{a} we assumed the same FWHM and $\Delta$v for all lines to fit these low S/N sources. \tablefoottext{b} The magnifications come from \citet{Spilker2016}. We have no good lens model for SPT0551-50 and we thus assume the median magnification of the SPT sample of 6.3 \citep{Spilker2016}. \tablefoottext{c} The average sample luminosity is determined combining detections and non-detections using the method described in Sects.\,\ref{sect:unbiased_ratio} and \ref{sect:deblend_ratio}. \tablefoottext{d}: The  S/N of the blend of HCN(5-4) and CN(4-3) is computed using the total flux of the two lines. \tablefoottext{e} The $^{13}$CO(4-3) line of SPT0125-47 is extracted assuming that the three lines have the same width and velocity offset as discussed in Sect.\,\ref{sect:line_ext}.}
\end{table*}

\subsection{Line extraction}

\label{sect:line_ext}

We extracted the lines by fitting the following model to the whole spectra. We assumed that the line profiles are Gaussian. We set a positivity prior on the line fluxes and allowed the full width at half maximum (FWHM) to vary between 200 and 1000\,km/s, which is typical for these types of sources \citep[e.g.,][]{Aravena2016}. We allowed a velocity offset up to 200\,km/s compared with the expected center of the line based on the systemic redshift estimated from the band-3 spectral scan of CO. Finally, we assumed a first-order baseline for the continuum emission of the sources. All the lines ($^{13}$CO(4-3), HCN(5-4), HCO$^{+}$(5-4), HNC(5-4), and CN(4-3)) and the continuum baseline are fitted simultaneously. This allows us to estimate the degeneracies between the baseline parameters and the line properties. In Appendix\,\ref{sect:fit_vs_mom0}, we explain why this method rather than the classical extraction in moment-zero maps was preferred.

We checked \textit{a posteriori} from the residuals that these are fair assumptions. We found no significant feature in the residual spectra. We computed the Pearson correlation coefficient between two neighboring channels and found that the correlation is lower than 0.2 for all our five residual spectra. Consequently, we can use these residuals to estimate the uncertainties on the line properties derived with our fit using a bootstrapping method. We thus took our best-fit model and added the residuals after randomly reshuffling the channels. Then, we refitted the result and saved the obtained best-fit parameters. We repeated this 10\,000 times. The error bars on the parameters are derived by computing the standard deviation of all these realizations. By construction, this method takes into account the degeneracies between parameters, and in particular between the baseline determination and the line fluxes. We estimated the signal-to-noise ratio (S/N) by dividing the best-fit peak flux density by its uncertainty. Because of the combined uncertainties on the peak flux density and the width, the relative uncertainty on the line flux is slightly higher than on the peak.

Only SPT0125-47 and SPT0551-50 have sufficiently bright lines to fit them directly with reasonable constraints on the FWHM and the velocity offset for each line. For the other sources, even if there is systematically a positive signal at the position of the lines, the S/N is often lower than 3 and the width and velocity of the lines cannot be constrained. We thus assume in our fit that these two properties are similar for all the lines. The low-frequency wing of the SPT0125-47 $^{13}$CO(4-3) line is not perfectly fitted, when the width and the velocity offset of each line are independent parameters. The FWHM is smaller than for other lines: 252$\pm$45 versus 475$\pm$128 and 496$\pm$119 for HCN(5-4) and HCO$^+$(5-4), respectively. However, this narrower profile is unlikely to be real (see Sect.\,\ref{sect:iso_profiles}) and might be caused by the noise. To determine the best fit of $^{13}$CO(4-3), we thus performed another fit assuming that the three lines have the same width and velocity offset. The flux is then slightly higher but consistent at 1\,$\sigma$ with the previous value: 1.15$\pm$0.22 versus 0.92$\pm$0.15\,Jy\,km/s.

The results of our line extraction are summarized in Table\,\ref{ref:lineext} and the best fit is shown in Fig.\,\ref{ref:spectra} (red solid line). When a line is not detected at $\geq$3\,$\sigma$, we derived a 3-$\sigma$ upper limit by summing the best-fit flux measured in the spectrum and three times the standard deviation of the flux in our multiple bootstrap realizations. Adding the signal present in the spectrum is crucial to obtain a reliable upper limit and is often forgotten in the literature. Otherwise, a source detected with an S/N of 2.9999 would have an $\sim$50\,\% probability of having a flux above the incorrectly computed 3-$\sigma$ upper limit.

When HNC(5-4) and CN(4-3) are both present in the spectra, a special method is used to extract them, since they are severely blended. The deblending of these two lines is discussed in Sect.\,\ref{sect:deblend_ratio}.

\subsection{Detected lines}

We detected three $>$5\,$\sigma$ lines in SPT0125-47. $^{13}$CO(4-3), HCN(5-4), and HCO$^+$(5-4) are detected at 7.2, 5.3, and 5.6\,$\sigma$, respectively. In SPT0551-50, there are $\sim$3\,$\sigma$ peaks at the position of the four targeted lines (3.4, 3.0, 3.1, and 2.6\,$\sigma$ for $^{13}$CO(4-3), HCN(5-4), HCO$^+$(5-4), and HNC(5-4), respectively). For the other three sources, only the blend HNC(5-4) and CN(4-3) in SPT0300-46 has an S/N larger than 3. However, these other data are very useful as lower or upper limits and allow us to derive unbiased mean properties for our sample using the method presented in Sects.\, \ref{sect:unbiased_ratio} and \ref{sect:deblend_ratio}. 

Previously, HCN(5-4), HCO$^{+}$(5-4), and HNC(5-4) have been detected at high
redshift only in quasars. \citet{Danielson2011} attempted to detect them in the
eyelash star-forming galaxy, but obtained only upper limits. The other
transitions of these three molecules have been found in star-forming galaxies
only below $z=2.5$. Our two detections in SPT0125-47 and our tentative detection
in SPT0551-50 push the high-z observations of these dense-gas tracers in
galaxies dominated by star formation to higher redshift. When we were finishing
this paper, we became aware that Chentao Yang et al. were also working on HCN
detections at z$\sim$3 in NCv1.143 and G09v1.97 and they kindly provided us with their
measurements (private communication\footnote{The PhD manuscript of \citet{Yang_PhD} can be downloaded
at \url{https://tel.archives-ouvertes.fr/tel-01661478/}.}).

The $^{13}$CO molecule has rarely been detected in star-forming galaxies at high
redshift. \citet{Danielson2013} found it in SMM J2135-0102 at $z=2.3$. In
addition, \citet{Weiss2013} reported two possible detections in the initial SPT
SMG redshift surveys: SPT0529-54 at $z= 3.36$, and SPT0532-50 at $z=3.39$. Our study doubles the number of $^{13}$CO detections at high redshift.


\begin{table*}
\caption{\label{table:ancillary_lines} Summary of the CO and [CI] data used in this paper. All these data, except the new detection of [CI](1-0) in SPT0125-47, are ancillary and were extracted from the redshift-search program \citep{Weiss2013} by \citet{Bothwell2016}.}
\centering
\begin{tabular}{lrrrrr}
\hline
\hline
Source  & SPT0103-45  & SPT0125-47& SPT0125-50  & SPT0300-46 & SPT0551-50\\ 
\hline
I$_{\rm ^{12}CO(4-3)}$ (Jy km/s)  & 11.7$\pm$0.7 & 23.1$\pm$0.6 & 7.9$\pm$1.0 & 4.9$\pm$0.5 & 12.0$\pm$0.8 \\ 
L'$_{\rm ^{12}CO(4-3)}$ (10$^{10}$ K km/s pc$^2$)  & 2.22$\pm$0.13 & 2.35$\pm$0.06\tablefootmark{a} & 1.20$\pm$0.15 & 1.22$\pm$0.13 & 3.39$\pm$0.24 \\
Data origin & \multicolumn{5}{c}{Data from z-search programs \citep{Weiss2013}, fluxes from \citet{Bothwell2016}} \\ 
 \hline 
I$_{\rm CI(1-0)}$ (Jy km/s)  &  --  & 6.3$\pm$0.2 & 2.4$\pm$0.5 & 1.8$\pm$0.8 &  --  \\ 
L'$_{\rm CI(1-0)}$ (intrinsic, 10$^{10}$ K km/s pc$^2$)  &  --  & 0.56$\pm$0.02 & 0.32$\pm$0.07 & 0.39$\pm$0.17 &  --  \\ 
Data origin &  no data & this paper & \multicolumn{2}{c}{z search \citep{Bothwell2016}} & no data \\
\hline
\end{tabular}
\tablefoot{\tablefoottext{a}{Converted from the CO(3-2) flux using the mean flux ratio measured by \citet{Spilker2014} by stacking.}}
\end{table*}

\begin{table*}
\caption{\label{table:ratios} Line flux ratios measured in our sources. The mean ratio is computed combining all the sources for which these two lines were observed (see the description of the method in Sects.\,\ref{sect:unbiased_ratio} and \ref{sect:deblend_ratio}).}
\centering
\begin{tabular}{lrrrrrr}
\hline
\hline
Line flux ratio  & SPT0103-45  & SPT0125-47& SPT0125-50  & SPT0300-46 & SPT0551-50 & Mean ratio \\ 
\hline
\vspace{0.2cm} 
I$_{\rm HCO^{+}(5-4)}$/I$_{\rm HCN(5-4)}$ & -  & 1.05$_{-0.24}^{+0.30}$ & -  & -  & 0.78$_{-0.28}^{+0.49}$ & 1.00$_{-0.19}^{+0.23}$ \\ 
 \vspace{0.1cm} 
I$_{\rm HNC(5-4)}$/I$_{\rm CN(4-3)}$ & -  & -  & -  & -  & -  & 1.60$_{-0.82}^{+1.74}$ \\ 
 \vspace{0.1cm} 
I$_{\rm HNC(5-4)}$/I$_{\rm HCN(5-4)}$ & -  & -  & -  & -  & -  & 1.03$_{-0.39}^{+0.59}$ \\ 
 \vspace{0.1cm} 
I$_{\rm ^{12}CO(4-3)}$/I$_{\rm ^{13}CO(4-3)}$ & >10.9 & 19.7$_{-3.1}^{+4.0}$ & > 23.1 & -  & 33.7$_{-10.9}^{+21.1}$ & 26.1$_{-3.5}^{+4.5}$ \\
 \vspace{0.1cm} 
I$_{\rm HCN(5-4)}$/I$_{\rm ^{12}CO(4-3)}$ & < 0.067 & 0.043$_{-0.009}^{+0.010}$ & < 0.055 & < 0.095 & 0.020$_{-0.008}^{+0.010}$ & 0.030$_{-0.005}^{+0.006}$ \\ 
\vspace{0.1cm} 
I$_{\rm HCN(5-4)}$/I$_{\rm [CI](1-0)}$ & -  & 0.158$_{-0.032}^{+0.038}$ & < 0.215 & < 1.374 & -  & 0.129$_{-0.026}^{+0.031}$\\
\hline
\end{tabular}
\tablefoot{For the ratios between two dense-gas tracers, we provided values only when the two lines are detected at more than 2\,$\sigma$. Otherwise, the ratio is often compatible at 2\,$\sigma$ with both zero and infinity. Concerning the $^{12}$CO/ $^{13}$CO, only SPT0125-50 is observed and not detected and we consequently derived a lower limit on this ratio, since $^{12}$CO is well detected. For the ratio between HCN(5-4) and CO(4-3) or [CI](1-0), we derived upper limits, since the denominator (CO or [CI]) is often detected.}
\end{table*}

\subsection{Unbiased average luminosities and line ratios}

\label{sect:unbiased_ratio}

\begin{figure*}
\centering
\begin{tabular}{lcr}
\includegraphics[width=8cm]{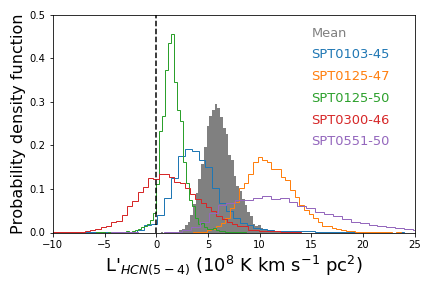} & \includegraphics[width=8cm]{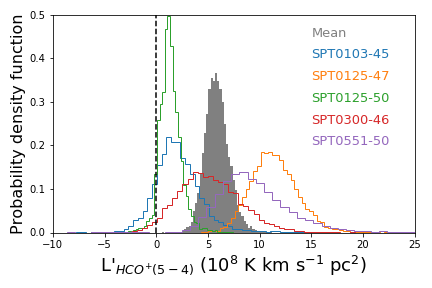} \\
\end{tabular}
\includegraphics[width=8cm]{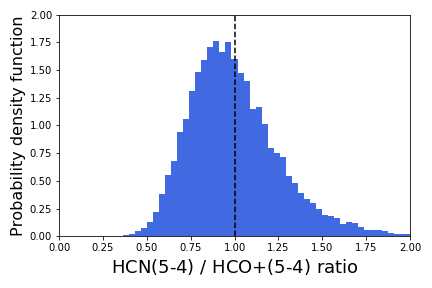}
\caption{\label{fig:pdf} Figures illustrating the method we used to derive unbiased average luminosity and line ratios (see Sect.\,\ref{sect:unbiased_ratio}). Upper left panel: Probability density distribution of the HCN(5-4) luminosity of our sources determined using a bootstrap technique (Sect.\,\ref{sect:line_ext}). The gray filled histogram is the probability distribution of the mean luminosity of our five sources (see Sect.\,\ref{sect:unbiased_ratio}). The vertical dashed line indicates the zero flux. Upper right panel: Same figure for HCO$^{+}$(5-4). Lower panel: Probability density distribution of the ratio between the mean HCN(5-4) and the mean HCO$^{+}$(5-4) luminosities. The vertical dashed line indicate the median value of the distribution.}
\end{figure*}

Given the relatively low line detection rate of our sample, we seek sample-averaged line luminosities and ratios without being biased towards the brightest sources. To this end, we used the bootstrap analysis results presented in Sect.\,\ref{sect:line_ext}. In Fig.\,\ref{fig:pdf}, we illustrate our method in the case of the average HCN(5-4) and HCO$^{+}$(5-4) luminosities and the flux ratio between these two lines. The probability density distribution (PDF) of the HCN(5-4) (upper left panels) and HCO$^{+}$(5-4) (upper right panel) luminosities of each source are represented as colored histograms. Concerning the $>$3\,$\sigma$ lines of SPT0125-47 and SPT0551-50, their PDFs exclude clearly the null hypothesis (L' = 0) as is expected for a (tentative) detection.

However, even for non-detections, the mode of the distribution is systematically above zero. Since these PDFs are quasi-Gaussian, the mode is very close to the best-fit value. This is the case for the eight $^{13}$CO(4-3), HCN(5-4), or HCO$^{+}$(5-4) lines present in our spectral windows. The case of HNC(5-4) and CN(4-3) is not discussed here but in Sect.\,\ref{sect:deblend_ratio}, since they are blended and the interpretation is more complicated. If the luminosity of these lines were strictly zero, the probability to have such a result would be (1/2)$^2$ = 0.3\,\%, since there would be a 50\,$\%$ chance for each individual PDF to have a negative mode. However, their flux is unlikely to be zero and could correspond to 1-3\,$\sigma$, since we planned our observation to attempt detections and the lines should not be too far from the detection threshold. Negative values of the mode of the distribution would thus be rather unlikely and would request a 1-3\,$\sigma$negative fluctuation of the noise in our observations. It is thus not so surprising that the modes of the PDFs of our non-detections are systematically positive. Indeed, the PDFs of such non-detections thus contain  weak but potentially useful information, if we combine several objects. However, of course, these lines cannot be qualified individually as detections, since their PDFs do not exclude negative values and our measurements are thus compatible with a zero luminosity.

We determined the average luminosity of the sample using both detections and non-detections by combining their PDFs. Since our observations of each source are independent, we can assume that the measured luminosities are independent. The PDF of the sum of the luminosities of our sources can then be computed by convolving the PDFs of their individual luminosities,
\begin{equation}
p \left (\sum_{k=1}^{N} L'_k = N \times \langle L' \rangle \right) = p_1(L'_1) \ast p_2(L'_2) \ast ... \ast p_N(L'_N),
\end{equation}
where $p_i(L'_i)$ is the PDF of the luminosity of the i-th source. The average is then computed by dividing this sum by N, the number of sources. In practice, we do not really need to compute this convolution analytically. We can just randomly draw luminosities from our 10\,000 bootstrap realizations for each of our sources and then sum them. We performed this operation 10\,000 times to obtain the PDF of the average luminosity of the sample. In Fig.\,\ref{fig:pdf}, the results are shown as gray filled histograms. As expected, the width of the PDF of the mean luminosity is significantly narrower than the PDFs of individual objects. We can also remark that the mean luminosity is clearly detected, since its PDF completely excludes zero.

We do not apply a positivity prior to perform our fits of the spectra. Indeed, in the hypothetical case of a line with a null flux for all sources, the bootstrap realizations would have a zero flux or a positive flux depending on the realization of the noise. The average flux would thus be positive even if the flux is zero for all sources. Practically, the absence of positivity prior would have a minor impact, since the probability of a negative flux is small ($<$10\%), except for HCN(5-4) in SPT0300-46 and $^{13}$CO(4-3) in SPT0125-50. Indeed, our unbiased mean luminosity estimates and the one derived with the positivity prior agree at better than 10\,\%, that is, 0.5\,$\sigma$.

This method is close to a stacking method, except that we first measure noisy fluxes and then average them instead of the contrary. However, since the lines of the various sources have different widths and  the baselines are difficult to subtract reliably, we preferred this approach to a standard stacking. Indeed, the stacked lines would have had a very peculiar shape, since these are the sum of lines with various widths. It would have been impossible to use our procedure to fit the lines and the baseline simultaneously. Finally, as explained in Appendix\,\ref{sect:fit_vs_mom0}, flux measurements based on stacked moment maps are also unreliable, since the baseline subtraction is potentially affected by biases due to the crowded spectral windows. A stacking procedure should not be used in this case, since the systematic biases could be similar for all objects and this bias would stay roughly constant with the number of sources, while the noise would decrease in 1/$\sqrt{N}$. The measurement would thus be dominated by systematic effects and not the instrumental noise.

We used a very similar method to derive mean line flux ratios. We first computed the PDFs of the mean of the line fluxes from our bootstrap realizations using the same method as previously described. However, to avoid giving more importance to low-z sources in the computation of the mean, we weighted the sources by the square of the luminosity distance. We then computed the ratio between these mean line fluxes.  The PDF of the ratio between two random variables is
\begin{equation}
p \left ( r = \frac{\langle L'_A \rangle}{\langle L'_B \rangle} \right ) \propto \int_{- \infty}^{+ \infty} |X| \, p_A(r X) \, p_B(X) \, dX,
\end{equation}
where $p_A(r X)$ and $p_B(X)$ are the PDF of the flux of the line A and B respectively, and $r$ is the line ratio in luminosity. In practice, the uncertainties on the ratio are determined by drawing realizations from the PDF of the mean luminosity of each line. The results are presented in Fig.\,\ref{fig:pdf}. The distribution is clearly asymmetric and we thus used the 16th and 84th percentile to produce the error bars in Table\,\ref{ref:lineext}. With this method, the contribution of the sources to the mean ratio is weighted by their luminosity. We tried to compute first the PDFs of the line flux ratios of individual sources and then average them, but this failed to get meaningful results. For low-S/N sources, the distribution is very broad with very large negative and positive outliers. Indeed, the PDF of the line flux in the denominator is compatible with zero and there are thus realizations for which the ratios are tending to $+\infty$ or $-\infty$.
In Appendix\,\ref{sect:average_simu}, we present the simulation used to validate this method.

\subsection{Deblending of HNC(5-4) and CN(4-3) and determination of unbiased mean ratios}

\label{sect:deblend_ratio}

HNC(5-4) is blended with CN(4-3). To our knowledge, the only previous individual detection of this blend at high redshift was performed by \citet{Guelin2007} in the lensed quasar APM 08279+5255 (see also \citealt{Riechers2007_CN} about the detection of CN(3-2) in the Cloverleaf). It is also detected in the stacked spectrum of all the SPT SMG sources observed by ALMA in cycle 0 \citep{Spilker2014}. \citet{Guelin2007} found that HNC(5-4) is 1.74 times brighter than CN(4-3) by fitting simultaneously the profile of the two lines in their spectrum. This last source has a different nature than ours, but it shows that, even if HNC(5-4) might dominate the flux, the CN(4-3) contamination cannot be neglected. It is thus important to deblend the two lines in order to put a constraint on HNC(5-4).

To do so, we refitted the data with a slightly different method than the one described in Sect.\,\ref{sect:line_ext}. We fitted simultaneously all the lines in the spectrum including HNC(5-4) and CN(4-3) and forced all the lines to have the same width and velocity offset (including the sources with good S/N). The width and the velocity of the blended lines are thus strongly constrained by the other lines without being completely fixed. Since CN(4-3) and HNC(5-4) are not at the same exact frequency, these additional constraints are sufficient to extract information from the data. We also have to apply a positivity prior on the flux of HNC(5-4) and CN(4-3), since the degeneracies between the two fluxes tend to produce negative values to overfit noise patterns. Even under these assumptions, the uncertainties on the ratio for an individual source remain very high, with relative uncertainties higher than 50\%. We thus derived the mean luminosities of the two lines and the mean line flux ratio between HNC and CN using the method presented in Sect.\,\ref{sect:unbiased_ratio}. We found a mean HNC(5-4)/CN(4-3) ratio of 1.60$_{-0.82}^{+1.74}$. This value is close to the value found  in the APM 08279+5255 quasar by \citet{Guelin2007}.

\subsection{CO and [CI] lines extracted from ancillary data}

\label{sect:ancillary_ext}

We want to compare our dense-gas tracers with lines from cold gas at lower density (CO, [CI]) found by our previous redshift search programs \citep{Weiss2013, Strandet2016}. The line fluxes of the [CI](1-0) line and the CO transitions covered by the redshift-search spectral scans were extracted in \citet{Bothwell2016}. For most of our sources, the CO(4-3) line falls in a frequency window covered by the redshift-search data. We thus chose to use this transition available for most of our sources as the reference one for CO in our analysis. For SPT0125-47, which is at lower redshift, only CO(3-2) is available. We derived the expected CO(4-3) flux using the line ratio measured in the stacked spectrum of the SPT SMG sources derived by \citet[][$I_{\rm CO(4-3)} / I_{\rm CO(3-2)} = 0.7$]{Spilker2014}. Finally, we detected [CI](1-0) in SPT0125-47 using our new ALMA data (see Appendix\,\ref{sect:ci}). The ancillary data used in this paper are summarized in Table\,\ref{table:ancillary_lines}.

\section{Dense-gas tracers: HCN, HCO$^{+}$, and HNC}

\label{sect:dense_gas_tracers}

\subsection{Scaling relation between the HCN and HCO$^{+}$ flux and the infrared luminosity}

\begin{figure}
\centering
\includegraphics[width=9cm]{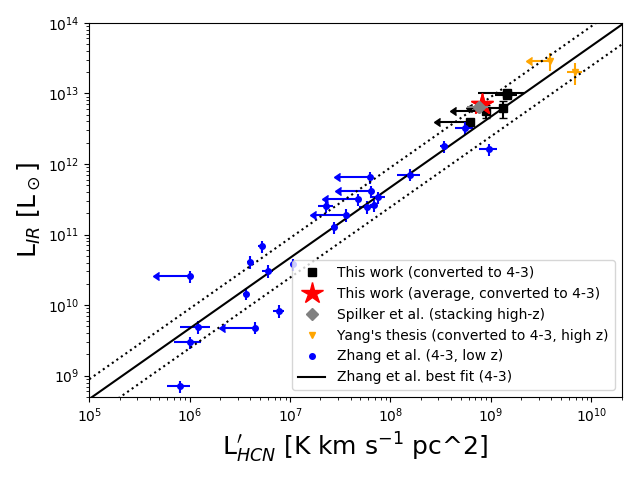}
\includegraphics[width=9cm]{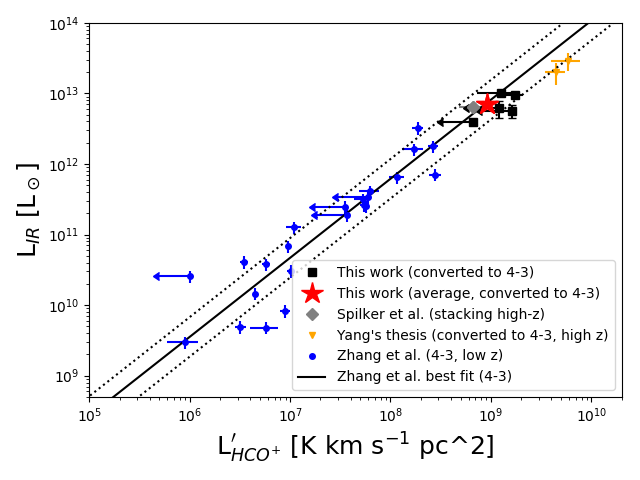} 
\caption{\label{fig:lir_scaling} Upper panel: Scaling relation between
the HCN(4-3) luminosity (converted from 5-4, see Sect.\,\ref{sect:lir_scaling})
and the total infrared luminosity. The black squares are the individual values
found in our high-z sample (after correcting for the lensing magnification). The
red star shows the average intrinsic L'$_{\rm HCN}$ and L$_{\rm IR}$ of our
sample. The gray diamond is the mean position of the SPT SMG sample derived from
the stacking of the ALMA cycle-0 data by \citet{Spilker2014}. The orange
downwards triangles are two H-ATLAS sources provided in Chentao Yang et al.
(private communication). The blue filled circles are from the local sample of
\citet{Zhang2014}. The black solid line is the best-fit relation of
\citet{Zhang2014} and the dotted lines indicate the 1-$\sigma$ intrinsic scatter around it. Lower panel: Same figure but for HCO$^{+}$.}
\end{figure}

\label{sect:lir_scaling}

The relation between the dense-gas content, traced by the HCN(1-0) line, and the SFR, traced by L$_{\rm IR}$, was found to be linear
in the local Universe by \citet{Gao2004}. 
A linear correlation between L$_{\rm IR}$ and
L'$_{\rm HCN(1-0)}$ is consistent with the simple physical picture in which the giant
molecular clouds (GMCs), traced by HCN, convert a fixed fraction of their mass
into stars before being disrupted
\citep[e.g.,][]{Faucher-Giguere2013,Grudic2018}. However, further studies found a
sublinear slope for the J=3-2 transition \citep{Bussmann2008,Juneau2009}, which
could have an impact on the interpretation of the physics of the star formation in
infrared-luminous objects and active galactic nuclei \citep[AGN, e.g.,][]{Narayanan2008}. More recently, after
considering careful aperture corrections, \citet{Zhang2014} found a linear slope
in the local Universe for the J=4-3 transition. The L$_{\rm IR}$-L'$_{\rm CO}$ relations
inferred for nearby galaxies have been found to be linear for 
$J=6-5$ \citep{Greve2014,Liu2015,Kamenetzky2016,Yang2017} 
and, possibly remain linear up to transitions as high as $J=12-11$ \citep{Liu2015}.
Extreme galaxies, however, such as the local ULIRG population and high-$z$ starbursts,
show sublinear L$_{\rm IR}$-L'$_{\rm CO}$ relations for $J=7-6$ and higher, due
to large amounts of energy being injected, likely via mechanical heating,
into a warm, dense, and non-star-forming ISM component \citep{Greve2014,Kamenetzky2016}.

At high redshift, the L$_{\rm IR}$-L'$_{\rm HCN}$ relation is
poorly constrained due to the small number of detections, all of which probe the high luminosity regime. However, \citet{Riechers2007} concluded, based
on the couple of detections and upper limits obtained at that time, that the
L$_{\rm IR}$/L'$_{HCN}$ ratio must be higher in high-z starbursts and quasars.
This trend agrees with the theoretical model of, for example, \citet{Krumholz2007},
which predicts a superlinear trend for HCN and HCO$^{+}$ at high infrared
luminosity. Similarly, \citet{narayanan11a} predicted a similar superlinear
trend for CO at very high L$_{\rm IR}$. According to them, the median density of galaxies
approaches the effective density of the molecular tracers, and the L$_{\rm
IR}$-L$_{\rm mol}$ relation will approach the L$_{\rm IR}$-M$_{\rm gas}$
relation, which in both models has a 1.5 exponent and is thus superlinear. On
average, in high L$_{\rm IR}$ systems such as DSFGs, the median gas density would be much
higher than in the lower luminosity systems investigated by \citet{Gao2004} for
which they found a linear trend.

To test if our galaxies follow the relation of \citet{Zhang2014}, we converted
the L'$_{\rm HCN(5-4)}$ of our objects into L'$_{\rm HCN(4-3)}$ assuming the
line ratio measured in the local Universe by \citet[][i.e., L'$_{\rm HCN(5-4)}$ /
L'$_{\rm HCN(4-3)}$ = 0.73]{Mills2013}, which is also compatible with the
measurements of \citet{Knudsen2007} in NGC253. We also corrected our sources for the magnification based on \citet{Spilker2016}. The results are presented in Fig.\,\ref{fig:lir_scaling} (upper panel). In addition to our detections and
upper limits (black squares), we also show the average L'$_{\rm
HCN(4-3)}$ and L$_{\rm IR}$ of our sample (red stars). We also compared the mean properties of our sample with the mean L'$_{\rm HCN(4-3)}$ measured by
\citet{Spilker2014} from a stacked spectrum of six SPT sources
in a slightly lower redshift range than our targets. To place their data in the diagram, we divided the mean L'$_{\rm
HCN(4-3)}$ and L$_{\rm IR}$ of their six sources by their mean
magnification, where a sample-median magnification of $\mu$=6.3 was assumed
for sources with unknown magnification factor. Our new measurements are compatible with the previous results
obtained by stacking.

All our objects have a small deficit of HCN luminosity compared to the
\citet{Zhang2014} relation (black line), perhaps similar to the deficit in HCN(2-1) relative to infrared (IR) 
observed in high-$z$ quasars by \citet{Riechers2007} (after accounting for the AGN contribution to the
IR luminosity). This explains \textit{a posteriori} why we did not reach
our initial goal of 5\,$\sigma$ detections in our four z$>$3 sources (see
Sect.\,\ref{sect:obs}). From the \citet{Zhang2014} relation, we expect a mean
L'$_{\rm HCN(5-4)}$ of 11.1$\times$10$^8$\,K\,km/s\,pc$^2$ (after converting to
the 5-4 transition) based on the infrared luminosity of our sources. We measured
an average flux of (7.1$\pm$1.6)$\times$10$^8$\,K\,km/s\,pc$^2$. This
corresponds to a deficit by a factor of 1.6 (0.20\,dex) corresponding to a
2.5\,$\sigma$ difference. However, \citet{Zhang2014} used the \citet{Sanders2003}
method to derive L$_{\rm IR}$, while another method was used for the SPT SMG
sample (\citealt{Blain2003}, model 1). We estimated the median factor between
the two methods by fitting the photometric points used by \citet{Zhang2014} with
the SPT SMG method. We found that their method derives luminosities that are a
factor of 1.27 higher. When we take this into account, the tension increases up
to a factor of 2.0 (0.30\,dex). In contrast, the HCN(5-4)
detection of NCv1.143 and the upper limit towards G09v1.97 (provided by C.~Yang)
are both, when converted to HCN(4-3), consistent with the observed local relation.

We checked if the deficit found with our sample could be explained by sample variance. The scatter
around the \citet{Zhang2014} relation is $\sim$0.28\,dex and we thus expect a
1-$\sigma$ sample variance for five objects of 0.28 / $\sqrt{5}$ = 0.125\,dex. If
we combine this with our measurement uncertainties, the significance of the HCN
deficit has thus only a $\sim$2\,$\sigma$ significance. It is therefore
not possible to draw any firm conclusions, especially considering the
uncertainties involved in converting our HCN(5-4) fluxes to HCN(4-3). Apart from
intrinsic scatter in the 5-4/4-3 ratio, systematic effects could also be
introduced by applying a locally-determined ratio to high-z starburst galaxies.
More data will be needed in the future.


We performed a similar analysis for the HCO$^{+}$. The conversion factor from
the 5-4 to the 4-3 transition based on \citet{Mills2013} is 0.64. Our results
agree with the \citet{Zhang2014}\footnote{There is a small mistake in
\citet{Zhang2014} and they provided us with an updated relation (private
communication), which we used in our analysis: log(L$_{\rm IR}$) = 1.12
log(L'$_{\rm HCO^{+}(5-4)}$) + 2.83.} relation (see Fig.\,\ref{fig:lir_scaling},
lower panel) and the average properties of our sources agree with the
measurements by stacking of \citet{Spilker2014}. Based on the \citet{Zhang2014}
relation and the mean infrared luminosity of our sample, we expect a mean
L'$_{\rm HCO^{+}(5-4)}$ of 6.9$\times$10$^8$\,K\,km/s\,pc$^2$ (after applying
the correction to homogenize the L$_{\rm IR}$, see above) and we found
(6.7$\pm$1.3)$\times$10$^8$\,K\,km/s\,pc$^2$, which is in excellent agreement.
Within the scatter, the HCO$^+$ detections provided by C.~Yang 
are also in agreement with the relation.

\begin{figure}
\centering
\includegraphics[width=9cm]{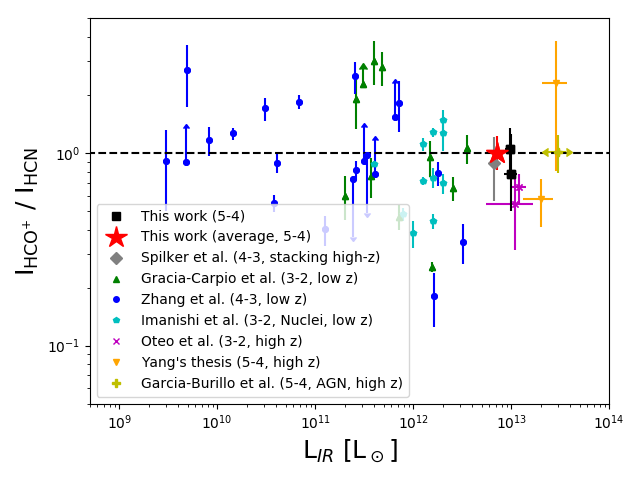}
\caption{\label{fig:hcop_hcn} Line flux ratio between HCO$^{+}$ and HCN as a function of the infrared luminosity. The black squares are the individual values obtained for our sources (only sources with two $>$3\,$\sigma$ lines are shown) and the red star is the average value of our five objects. The gray diamond is the mean ratio found by \citet{Spilker2014} using a stacking of the cycle-0 SPT SMG spectra. The green triangles are the ratios of the 3-2 transitions in the local sample of \citet{Gracia-Carpio2008} and blue filled circles are the 4-3 ratios in the low-z sample of \citet{Zhang2014}. The cyan pentagons represent the 3-2 ratios measured in local active nuclei by \citet{Imanishi2016}. The purple crosses shows the 3-2 ratio measured in z$\sim$1.5 lensed star-forming galaxies by \citet{Oteo2016}. The orange downwards-facing triangles are two H-ATLAS sources provided by Chentao Yang et al. (private communication). The yellow plus is the 5-4 flux ratio found by \citet{Garcia-Burillo2006} in the APM08279+5255 quasar (see also \citealt{Riechers2010} for the 6-5 transition).}
\end{figure}

\subsection{The HCO$^{+}$ to HCN flux ratio}

We compared the HCO$^{+}$/HCN $J=5-4$ flux ratios of our SPT sources (see Table\,\ref{table:ratios}) with HCO$^+$/HCN ratios
of low- and high-$z$ galaxies from the literature (Fig.\,\ref{fig:hcop_hcn}). The main advantage of
using flux ratios is that they cancel out the magnification factor $\mu$ if the
differential magnification \citep{Hezaveh2012,Serjeant2012} is negligible (see
Sect.\,\ref{sect:profile_comp}). While some of the HCO$^+$/HCN ratios from the literature were
for transitions other than $J=5-4$, we proceeded to compare them with our ratios under the assumption
that the spectral line energy distributions (SLEDs) of the HCO$^{+}$ and HCN are similar. In addition to the average value ($1.00_{-0.19}^{+0.23}$) of our
sample, we put in our diagram only the sources that have at least a
3\,$\sigma$ signal at the position of each line (SPT0125-47 and SPT0551-50). For the other sources with $<$3\,$\sigma$ signal, both the numerator (HCO$^{+}$ flux) and the denominator (HCN flux) are compatible with zero at 3\,$\sigma$ (see Sect.\,\ref{sect:unbiased_ratio}). The PDF of their ratio is thus compatible with both zero and infinity at 3\,$\sigma$ and deriving upper or lower limits does not make sense. Nevertheless, we checked the average line ratio derived for these three other sources and found
that it is compatible at 1\,$\sigma$ with the average value derived for the fives sources and
the two individual measurements, but with very large uncertainties (0.98$_{-0.40}^{+0.73}$).

Similar to the local star-forming samples of \citet{Gracia-Carpio2008} and
\citet{Zhang2014}, the mean ratio of our sample is compatible with unity. This
is consistent with the ratio derived for the J=4-3 transitions using the stacked
spectra of \citet{Spilker2014}, but with improved uncertainties. The mean ratio
of our sample is 2\,$\sigma$ higher than in the two high-z star-forming galaxies
of \citet{Oteo2016}. NCv1.143 (Yang et al.) has similar values to these two
objects, but G09v1.97 has a much higher value ($\sim$2). However, all these high-z sources are in the intrinsic scatter of the local relation.
As in the low-z Universe, the ratio varies significantly within the population of lensed
DSFGs. \citet{Braine2017} found that low-metallicity regions of local galaxies
($<$0.5\,Z$_\odot$) have a high HCO$^+$/HCN flux ratio ($\sim$2) instead of a
ratio close to unity. Finding a unity ratio in our sources could be consistent 
with an already mature ISM at early cosmic times, but a larger statistical sample will be necessary to confirm or not this possibility.

Recently, \citet{Imanishi2016} and \citet{Izumi2016} found a  HCO$^+$/HCN
flux ratio ($\sim$0.5) for both the $J=3-2$ and $J=4-3$ transitions that is lower in local AGNs than in star-forming galaxies\footnote{In
the original articles, the authors used the HCN/HCO$^+$ flux ratio (the inverse)
and thus discussed a high ratio instead of a low ratio in this paper.} and
proposed that this quantity could be used to identify AGNs. \citet{Izumi2016}
discussed two explanations for the low HCO$^{+}$/HCN: an enhanced abundance of HCN compared with HCO$^{+}$
coming from the complex chemical and radiative mechanisms involving these
molecules in the neighborhood of an AGN or a systematically higher gas density
around AGNs. We found a ratio close to unity, which is consistent
with our objects being star-formation dominated. However, we should interpret this simple
diagnostic with caution, since some AGNs of \citet{Imanishi2016} and the
APM08279+5255 quasar at $z = 3.9$ have flux ratios close to unity.

\begin{figure}
\centering
\includegraphics[width=9cm]{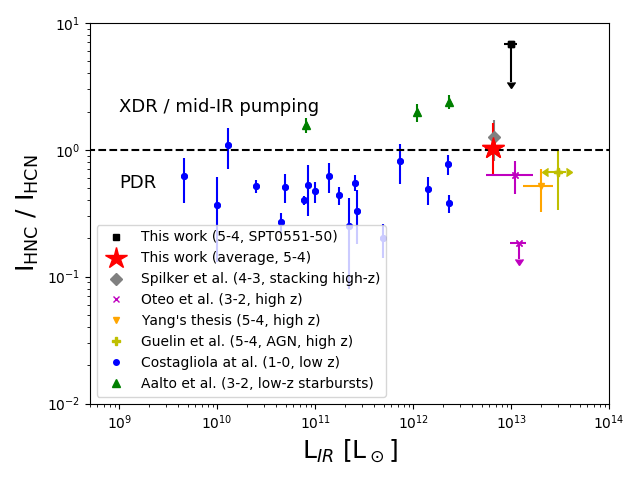} \\
\caption{\label{fig:hnc_hcn} Line flux ratio between HNC and HCN as a function of the infrared luminosity. The red star is the mean ratio found in our sources and the gray diamond is the mean ratio found by \citet{Spilker2014} using a stacking of the cycle-0 SPT SMG spectra. The black square is the upper limit determined for SPT0551-50. For the other sources, the individual ratio cannot be constrained, since both lines are too weak to derive an upper or a lower limit. The blue filled circles represent the local sample of \citet[][1-0 transition]{Costagliola2011}. The green triangles are the 3-2 transitions in the Arp220, Mrk231, and NGC4418 starbursts \citep{Aalto2007}. The purple crosses are the two z$\sim$1.5 lensed star-forming galaxies of \citet{Oteo2016} and the yellow plus is the APM08279+5255 quasar \citep[][see also \citealt{Riechers2010} for the 6-5 transition]{Guelin2007}. The orange downwards triangles are  two H-ATLAS sources provided in Chentao Yang's thesis (private communication).}
\end{figure}

\subsection{The HNC to HCN flux ratio}

Other important diagnostics can be performed using the isomer ratio between HNC
and HCN. HNC traces gas of similar density to HCN, but the observed HNC/HCN flux ratio is close to unity in dark clouds and up to 100 times smaller in hot
environments \citep{Schilke1992,Hirota1998}. In addition, \citet{Aalto2007}
showed that the HNC/HCN flux ratio can be above unity in local starbursts, which
is not intuitive because HNC should be more easily destroyed than HCN by the
strong radiation fields and high temperatures in these objects. They proposed
two possible explanations: HNC is excited by mid-IR pumping of its rotational
levels or X-ray dissociation regions (XDRs) have an impact on the abundance of
HNC. Since then, the case of Arp220 has been extensively investigated and more
signs of HCN or HNC pumping have been identified
\citep[e.g.,][]{Aalto2015,Galametz2016}. Finally, a similar scenario was also
discussed in \citet{Weiss2007} to explain HCN luminosity in the high-redshift
quasar APM08279+5255.

Because of the blending of HNC(5-4) with CN(4-3), we derived only a mean HNC/HCN ratio using the method described in Sect.\,\ref{sect:deblend_ratio}, and found a ratio compatible with unity (1.03$_{-0.39}^{+0.59}$). This ratio is compatible with the one obtained by stacking of the HNC(4-3) and HCN(4-3) lines by \citet[][gray diamond in Fig.\,\ref{fig:hnc_hcn}]{Spilker2014}. Since the HCN(4-3) line is not blended, it is thus reassuring to find similar values. Our sources are at the border between the regime dominated by photodissociation regions (PDRs) and the domain, where XDR and/or mid-IR pumping are necessary to explain the line ratios. In Fig.\,\ref{fig:hnc_hcn}, we compare the mean ratio found for our sample (derived using the method described in Sect.\,\ref{sect:deblend_ratio}) with local and distant samples. The mean HNC/HCN ratio of our sample is a factor of approximately three above the local IRAM/30m sample of \citet{Costagliola2011}, but lower by a factor of approximately two than the starbursts of \citet{Aalto2007}. Our average measurements are only 1\,$\sigma$ above the measurements of APM08279+5255 by \citet{Guelin2007}, SDP.9 of \citet{Oteo2017}, and NCv1.143 in Yang's thesis, but an order of magnitude above the upper limit for SDP.11 \citep{Oteo2017}.

\begin{figure*}
\centering
\begin{tabular}{cc}
\includegraphics[width=9cm]{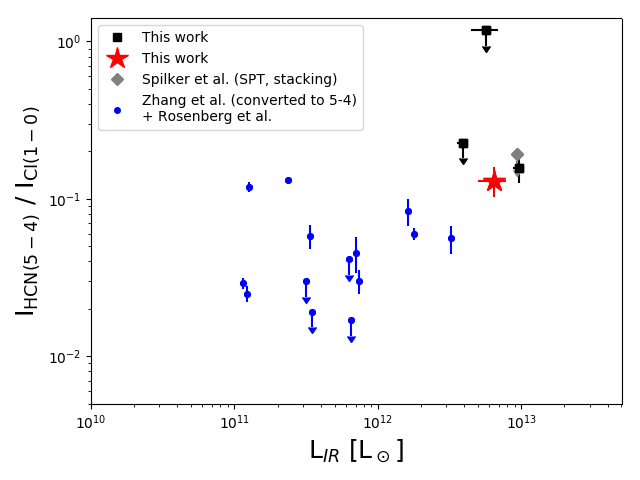} & \includegraphics[width=9cm]{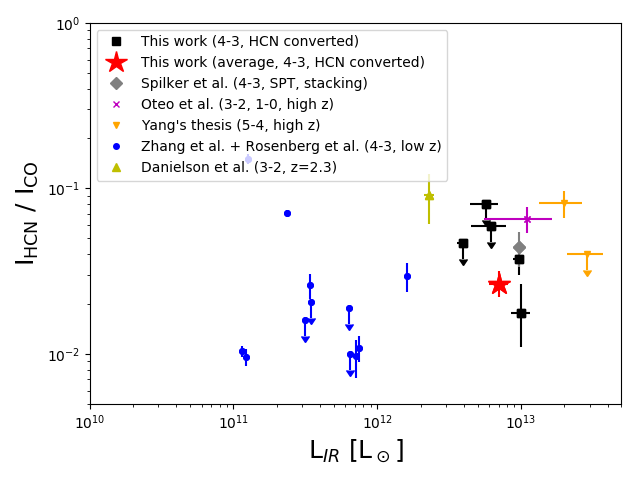} \\
\end{tabular}
\caption{\label{fig:cociratios} Left panel: Flux ratio between HCN(5-4)
and [CI](1-0) as a function of the infrared luminosity. Our sources are
represented by black filled squares and the average by a red star. The gray
diamond represents the stacking of the cycle-0 SPT SMG spectra
\citep{Spilker2014}. The blue filled circles are the local HCN(4-3) measurements
of \citet{Zhang2014}, where HCN(4-3) is converted into HCN(5-4) following the
recipe described in Sect.\,\ref{sect:lir_scaling}, combined with the [CI](1-0)
measurements of \citet{Rosenberg2015}. Right panel: Ratio between
HCN(4-3) and $^{12}$CO(4-3) as a function of the infrared luminosity. As
explained in Sect.\,\ref{sect:dgf}, we used the HCN(5-4)/CO(5-4) or
HCN(3-2)/CO(3-2) ratios, when the 4-3 transitions were not available. The purple
cross is the measurement of \citet{Oteo2016} in SDP.9. The orange downwards
triangles are two H-ATLAS sources provided in Chentao Yang's thesis (private
communication). The yellow triangle is the measurements of \citet{Danielson2011}
in SMM J2135-0102.}
\end{figure*}


\section{Dense gas versus lower density tracers (CO, [CI])}

\label{sect:comp_gas_tracers}

The origin of the strong star formation in the most extreme high-redshift
starbursts is a source of intense debates
\citep[e.g.,][]{Engel2010,Daddi2010b,Hayward2011,Hayward2013a,Carilli2013,Casey2014,Narayanan2015}.
Large gas reservoirs are not sufficient to explain the SFR of the most extreme
systems and a temporary increase of the star formation efficiency, measured relative to
the total gas mass as traced by CO(1-0) or CO(2-1), is
necessary \citep[e.g.,][]{Daddi2010b,Genzel2010,Sargent2014}. \citet{Gao2007}
and \citet{Daddi2010b} suggested that the increase of the star formation efficiency in these objects
is linked to an increase of the dense-gas fraction (DGF). The dense gas fraction
is thus one of the keys to understanding the nature of this type of sources. We note, however, that an increase in the dense-gas star formation efficiency is not required in this picture.

\subsection{Dense gas fraction versus infrared luminosity}

\label{sect:dgf}

In this section we compare the flux ratio between HCN(5-4), which probes the dense gas, and [CI](1-0), which is
thought to be a tracer of the bulk gas reservoir. For the three sources in our sample with [CI](1-0)
measurements, we therefore adopt this ratio as a proxy for the DGF.

To build a reference sample in the local Universe, we combined the HCN(4-3)
sample of \citet{Zhang2014} with the [CI] fluxes from the local
\textit{Herschel}/spectral and photometric imaging receiver (SPIRE) spectroscopic sample of \citet{Rosenberg2015}. We adopted the aperture-corrected integrated [CI] fluxes published by \citet{Rosenberg2015}, and, similarly, the aperture-corrected integrated HCN(4-3) fluxes published by \citet{Zhang2014}. Finally, the HCN(4-3) fluxes were converted into HCN(5-4) using the conversion described in Sect.\,\ref{sect:lir_scaling}.

In the left panel of Fig.\,\ref{fig:cociratios}, we show the ratio between
HCN(5-4) and [CI](1-0) versus L$_{\rm IR}$ for our three sources observed in [C{\sc i}](1-0).
Also shown is their mean flux ratio, as well as the upper limit derived from the stacked spectra of SPT sources 
\citep{Spilker2014}. Our sources are seen to be consistent with the stacked SPT value, but at the
limit of the upper envelope of the local reference sample.

Given the lack of low-$J$ CO transitions for some of our sources, we 
are not able to gauge the dense gas fraction using HCN(mid-$J$)/CO(low-$J$) ratio.
Instead we opted for the $J=4-3$ CO transition, since it is available for most of our
sources, and examine the HCN(4-3)/CO(4-3) ratio. HCN(4-3) and CO(4-3) 
have similar upper level energies ($E_J/k_{\rm B} \sim 40-55\,{\rm K}$) and both
trace dense gas, albeit HCN(4-3) has a $\sim 400\times$ higher critical density than CO(4-3)
($\sim 8\times 10^6\,{\rm cm^{-3}}$ vs. $\sim 10^4\,{\rm cm^{-3}}$). For the local reference sample we again adopt \citet{Zhang2014}, and use their
directly measured HCN(4-3) fluxes and the CO(4-3) fluxes from \citet{Rosenberg2015}
to form the HCN(4-3)/CO(4-3) ratios. 
In the right panel of Fig.\,\ref{fig:cociratios}, we compare the HCN(4-3)/CO(4-3) 
flux ratios of our sources, where we have converted HCN(5-4) to HCN(4-3).
There are not a lot of HCN(4-3)/CO(4-3) measurements of high-$z$ sources 
available in the literature, and we therefore extended our comparison to HCN(5-4)/CO(5-4)
and HCN(3-2)/CO(3-2) in order to facilitate a comparison with other high-$z$ starbursts. 
Our mean HCN(4-3)/CO(4-3) ratio is in the scatter of the local
values and compatible with the upper limit on G09v1.97 (Yang's thesis). In contrast,
our sources are on average a factor of 1.6 lower than the stacking measurement
of \citet{Spilker2014} on the full cycle-0 SPT SMG sample ($\sim$2\,$\sigma$
difference) and a factor of approximately three below the high-z measurements of
\citet{Oteo2016} in SDP.9, \citet{Danielson2011} in SMM J2135-0102, and NCv1.143
in Yang's thesis. 


\begin{figure}
\centering
\includegraphics[width=9cm]{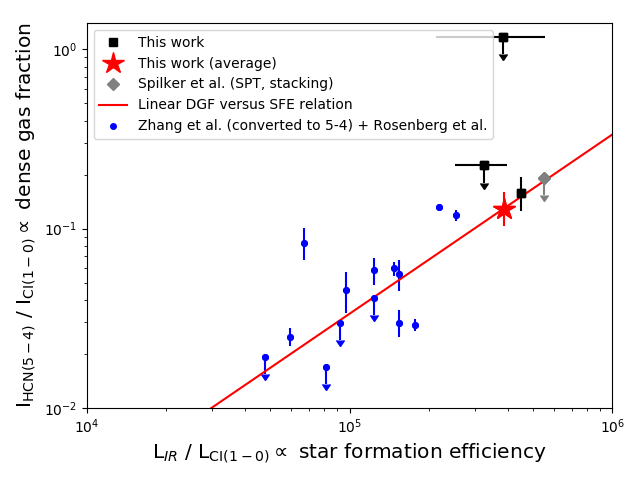}
\caption{\label{fig:sfe} Ratio between HCN(5-4) and [CI](1-0), tracing the dense-gas fraction, as a function of the ratio between the infrared and the [CI](1-0)
luminosity, tracing the star formation efficiency. The data comes from the same
references as in Fig.\,\ref{fig:cociratios}. The red solid line is a linear
relation between the dense-gas fraction and the star formation efficiency (DGF
$\propto$ SFE), which normalization has been set to match the mean value of our
sample (red star).}
\end{figure}

\subsection{A link between dense-gas fraction and star formation efficiency}

\label{sect:dgf_sfe}

\citet{Gao2007} found a correlation between L'$_{\rm
HCN(1-0)}$ / L'$_{\rm CO(1-0)}$ and L$_{\rm IR}$ / L'$_{\rm CO(1-0)}$, which 
they interpreted as a correlation between the DGF and the SFE (with respect to
the total gas mass). In Fig.\,\ref{fig:sfe}, we performed a similar diagnostic
using [CI](1-0) instead of CO(1-0) and HCN(5-4) instead of HCN(1-0). Our SPT
sources are consistent with the trend of DGF versus SFE found in our local
reference sample described in Sect.\,\ref{sect:dgf}. A similar result is found
if we use CO(4-3) instead of [CI](1-0) (see Fig.\,\ref{fig:sfe_co} in appendix).
This seems to indicate that the correlation between the DGF and the SFE is still
valid for the most star-forming systems at high redshift and for the densest gas
probed by the high-J transitions of HCN. However, an artificial correlation can
appear in diagrams representing A/C versus B/C. In Appendix\,\ref{sect:sanity},
we confirm that it is not the case for our analysis using two different total
gas mass tracers in the x and y axis.

The relation linking our local reference sample and the SPT sources suggests
that DGF $\propto$ SFE (solid red line in
Fig.\,\ref{fig:sfe}). This result agrees with the suggestion by \citet{Gao2007}
and \citet{Daddi2010b} that the high SFEs in starbursts are directly connected
to their DGF. The link between these two quantities is not surprising, if, as
suggested by the various physical models cited previously, the amount of dense
gas, rather than the total gas reservoir, drives the SFR. Unfortunately, our
observations do not allow us to form a conclusion about the mechanism that causes these
high DGFs. The sources in our sample are at the high end of the SFE and DGF
distribution of our local reference sample. Their impressive SFRs are thus
caused by a combination of large gas reservoirs \citep{Aravena2016} coupled with
a high DGF.

\begin{figure}
\centering
\includegraphics[width=9cm]{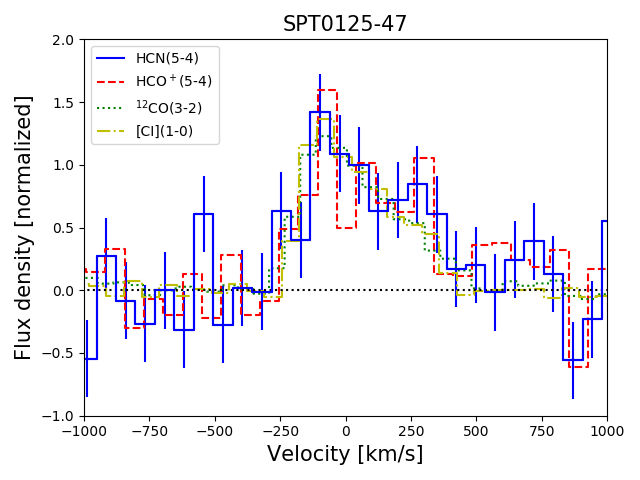}
\caption{\label{fig:profile_comp} Comparison of the velocity profile of the dense-gas lines (HCN(5-4) in blue solid line and HCO$^{+}$(5-4) in red dashed line, and other gas tracers (CO(3-2) in green dotted line and [CI](1-0) in yellow dot-dash line. We plotted the channel RMS uncertainties of HCN(5-4). HCO$^{+}$(5-4) has similar channel uncertainties (not plotted). $^{12}$CO and [CI] have a much better S/N and their uncertainties can be neglected. We normalized all the lines to have $\int S_{\nu} \, dv$ = 500 Jy\,km/s. With this normalization, the peak flux of a rectangular line with a typical 500\,km/s width is unity.} 
\end{figure}

\subsection{Similarity of the line profiles in SPT0125-47 and differential lensing}

\label{sect:profile_comp}

In SPT0125-47, dense-gas tracers are detected at more than 5\,$\sigma$ and it is
thus possible to compare the line profile of HCN and HCO$^{+}$ with [CI] and CO
(see Fig.\,\ref{fig:profile_comp}). Our high-S/N [CI] detection and the
ancillary CO(3-2) line detection from the redshift search have a similar
asymmetric profile with a much broader redshifted tail. A similar asymmetry is
found for HCN(5-4) and HCO$^{+}$(5-4). This suggests that molecular regions from
the lowest to the highest density are distributed in the same way across this
object. 

The similarity of the profiles also suggests that
differential lensing \citep{Hezaveh2012,Serjeant2012} should not be too strong
in this object. Concerning the other objects of our sample, we do not have a sufficient S/N to be able to check the similarity of the profiles between the $^{12}$CO lines and the dense-gas lines. However, as discussed in \citet{Gullberg2015}, most of the SPT sources have for instance similar CO and [CII] profiles. Except if the dense-gas lines have a fundamentally different spatial distribution than lower density tracers, we have no good reason to expect \textit{a priori} any strong differential lensing effect. We thus neglected this effect in this paper. The only way to measure the impact of the differential lensing on integrated fluxes would be to perform high-resolution imaging of dense-gas lines, but it is out of reach of ALMA in a reasonable amount of time.

\begin{figure}
\centering
\includegraphics[width=9cm]{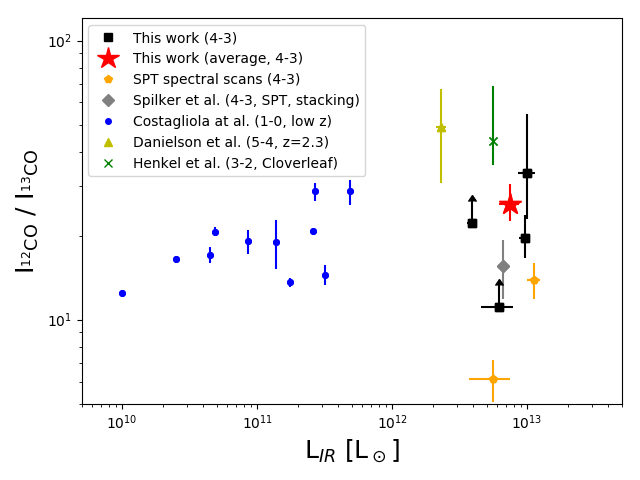}
\caption{\label{fig:isotopes} Flux ratio between the $^{13}$CO and $^{12}$CO as a function of the infrared luminosity. The same transition of both lines is used to compute these ratios. Our sources are represented by black filled squares and their average properties by a red star. The two serendipitous detections obtained during the first redshift-search campaign in SPT0529-54 and SPT0532-50 \citep{Weiss2013} are shown using orange pentagons. The gray diamond shows the average ratio measured by \citet{Spilker2014} using a stacking of the full redshift-search SPT sample. The local sample from \citet{Costagliola2011} is plotted with a blue filled circle. The yellow triangle and the green cross are the high-z measurements of \citet{Danielson2013} in SMM J2135-0102 and \citet{Henkel2010} in the Cloverleaf quasar.}
\end{figure}

\begin{figure}
\centering
\includegraphics[width=9cm]{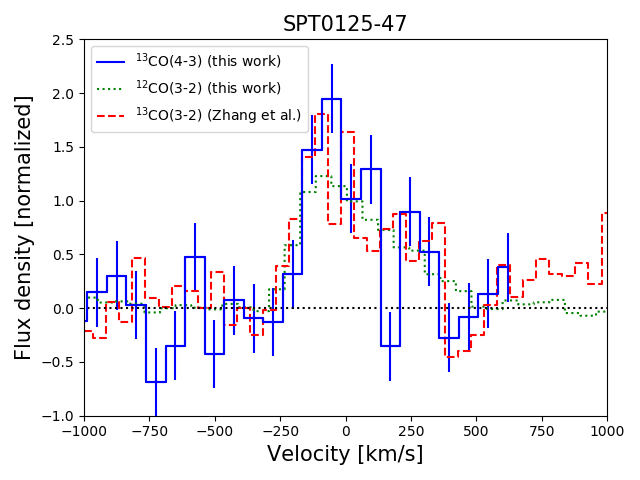} \\
\caption{\label{fig:isotope_comp} Comparison of the velocity profiles of the $^{13}$CO(4-3) (blue solid line) and $^{12}$CO(3-2) lines (green dotted line). The uncertainties on the $^{12}$CO spectrum are not plotted, since they are negligible compared with the $^{13}$CO ones. We normalized all the lines to have $\int S_{\nu} \, dv$ = 500 Jy\,km/s. With this normalization, the peak flux of a line with a typical 500\,km/s width is unity. The red dashed line is the $^{13}$CO(3-2) profile measured by \citet{Zhang2018}.}
\end{figure}

\section{Properties of the CO isotopic lines}

\label{sect:isotopes}

\subsection{Flux ratios between $^{13}$CO and $^{12}$CO}

As explained in the Introduction, the [$^{12}$C]/[$^{13}$C] abundance ratio has been proposed as a diagnostic of the evolutionary stage of a galaxy, since different nuclear reactions produce $^{12}$C and $^{13}$C; the former is produced via triple alpha nuclear processes in young massive stars, while the latter is produced in the CNO cycle in evolved asymptotic giant branch (AGB) stars \citep{WilsonRood1994}.

A high [$^{12}$C]/[$^{13}$C] abundance ratio, therefore, could indicate a chemically young and largely unprocessed ISM \citep{Hughes2008,Henkel2010}. However, other physical mechanisms could result in a high [$^{12}$C]/[$^{13}$C] ratio, namely star formation from a top-heavy IMF in young starbursts \citep[e.g.,][]{Romano2017, Zhang2018} -- with the latter being the result of extreme cosmic-ray-dominated, star formation regions rather than a metal-poor gas. In intense starburst environments, the cosmic ray heating may be so severe that dense, deeply embedded molecular cores are significantly heated, resulting in a raise of the Jeans mass floor \citep{Papadopoulos2011}.

The [$^{12}$C]/[$^{13}$C] ratio is typically constrained indirectly from observations of CO, HCN, and HCO$^{+}$ and their $^{13}$C isotopologues. This approach, however, is complicated by isotope fractionation where chemical reactions involving $^{13}$C are energetically favored over the same reactions involving $^{12}$C \citep{Watson1976,Langer1984}. The expectation from isotope fractionation is that [$^{12}$CO]/[$^{13}$CO] provides a lower limit to [$^{12}$C]/[$^{13}$C] \citep{Langer1984,Tunnard2016}.

Furthermore, in intense far-UV environments, selective photodissociation can lead to an increase in [$^{12}$CO]/[$^{13}$CO], since $^{13}$CO is readily destroyed given its low abundance, while $^{12}$CO will be able to self-shield \citep{Bally1982}. However, this effect is expected to play a role only in diffuse gas (A$_{\rm V} \sim$1) and it is thus doubtful whether selective photodissociation plays a significant role in dusty galaxies, where FUV light undergoes heavy extinction \citep{Casoli1992, Papadopoulos2014}.

Finally, it is non-trivial to infer $^{12}$CO-to-$^{13}$CO abundance ratios from their line intensity ratios. This is because the $^{12}$CO lines are optically thick, while the rarer $^{13}$CO lines tend to be optically thin \citep[e.g.,][]{Casoli1992,Aalto1995}. Optical depth effects can thus complicate the picture, since in the case where $^{12}$CO is optically thick the $^{12}$CO/$^{13}$CO line intensity ratio will be an upper limit to the [$^{12}$CO]/[$^{13}$CO] abundance ratio.

In Fig.\,\ref{fig:isotopes} (left), we compare our measurements of the $^{12}$CO versus $^{13}$CO ratio with the literature. We found a mean ratio of  26.1$_{-3.5}^{+4.5}$, which is two times higher than the measurements of \citet{Spilker2014} using the stacked spectra of all SPT SMGs observed by ALMA during the cycle 0. Our sample was selected because of their high apparent luminosity and could be biased towards objects with a stronger contribution of the recent star formation to the ISM enrichment. It is therefore not surprising to find a lower $^{13}$CO abundance, and thus a higher $^{12}$CO/$^{13}$CO ratio, in our subsample. \citet{Weiss2013} reported a detection of $^{13}$CO(4-3) in SPT0529-54 and a tentative detection in SPT0532-50 in the initial SPT SMG redshift-search program. The fluxes of these two lines, not published in the original paper, are 1.36$\pm$0.24\,Jy\,km/s and 1.07$\pm$0.32\,Jy\,km/s , respectively, while the $^{12}$CO(4-3) fluxes are 8.32$\pm$0.34\,Jy\,km/s and 14.89$\pm$0.41\,Jy\,km/s. These two sources have much lower ratios than the new objects presented in this paper. This is not surprising, since these serendipitous detections are biased towards sources with particularly bright $^{13}$CO(4-3) lines or with a positive fluctuation of the noise at the position of this line, while the $^{12}$CO lines are brighter and thus systematically detected. Consequently, the $^{12}$CO/$^{13}$CO ratios of these serendipitous detections are biased toward lower values. In Appendix\,\ref{sect:bias_serendipitous}, we present a simulation illustrating this effect.

Our mean ratio corresponds to the upper envelope of the local sample of \citet{Costagliola2011} in the LIRG regime, but is lower than previous high-z measurements on individual objects of \citet[][Cloverleaf]{Henkel2010} and \citet[][SMM J2135-0102]{Danielson2013}. Despite our sources having a deficit of $^{13}$CO compared to the average values in the local Universe, their isotopic flux ratio is compatible with some local LIRGs. However, it is hard to know if it is really due to similar $^{13}$CO abundances in local LIRGs and DSFGs or if the $^{13}$CO abundance evolves and is compensated by optical depth effects. The presence of $^{13}$CO shows that our objects are clearly not experiencing a first giant starburst fed by pristine gas.


\subsection{Comparison of $^{13}$CO and $^{12}$CO line profiles in SPT0125-47}

\label{sect:iso_profiles}

If the various parts of a galaxy have very different ages and ISM enrichment, we might observe very different $^{12}$CO and $^{13}$CO profiles. In particular, we expect a stronger enrichment of $^{13}$CO in the center of galaxies, which harbor older stellar populations responsible for producing this isotope. Furthermore, levels of accretion of pristine cosmological gas onto the galaxy are lower in the central regions, resulting in less dilution of the enriched gas already there. In SPT0125-47, however, we found very similar $^{12}$CO and $^{13}$CO line profiles, suggesting the abundance ratio of the two molecules remains roughly constant throughout.

In Fig.\,\ref{fig:isotope_comp}, we compare the velocity profiles of $^{13}$CO(4-3) and $^{12}$CO(3-2). In addition, we included the $^{13}$CO(3-2) measurement of \citet{Zhang2018}. Overall, the $^{12}$CO and $^{13}$CO profiles are rather similar. The two channels at the peak of $^{13}$CO(4-3) are 1\,$\sigma$ and 2\,$\sigma$ above the $^{12}$CO(3-2), but this is not the case for $^{12}$CO(3-2). This small tension thus seems to be caused by noise. The similarity of the three profiles suggest that the maturity of the ISM across the different regions of SPT0125-47 is relatively homogeneous. These similar profiles are also a reassuring indication that there should be no strong differential magnification affecting this source.

\section{Conclusion}

In this paper, we presented the main results of our ALMA program targeting
dense-gas lines (HCN(5-4), HCO$^+$(5-4), and HNC(5-4)) in five strongly-lensed
DSFGs at 2.5$<z<$4. We obtained in total two detections (S/N>5) in SPT0125-47 and four tentative detections (S/N$\sim$3) of dense-gas lines in other sources. In addition, our observations yielded two detections (one of which is tentative), as well as two upper limits, of $^{13}$CO(4-3). We also detected [CI](1-0) with an S/N of 36 in SPT0125-47. Finally, we developed a
method to derive unbiased average properties of our sample to compare them with
the low- and high-redshift literature. From this information, we draw the
following conclusions:
\begin{itemize}
\item The average HCN(5-4) luminosity of our sample is formally a factor
of $\sim 1.7$ lower than what is expected from the local L'$_{\rm
HCN}$-L$_{\rm IR}$ linear relation of \citet{Zhang2014}. However, if we take
into account the sample variance, this tension decreases to 2\,$\sigma$. Furthermore,
we cannot rule out the introduction of systematic uncertainties in the conversion
from HCN(5-4) to HCN(4-3) when comparing to the \citet{Zhang2014} sample.
A similar trend, however, was reported in high-$z$ quasars for high-$J$ transitions by
\citet{Riechers2007} and is also expected by such theoretical models as
that of \citet{Krumholz2007} for high-SFR systems.
\item Our sample has a mean I$_{\rm HCO^{+}(5-4)}$/I$_{\rm HCN(5-4)}$ flux ratio of $1.00_{-0.19}^{+0.23}$. This ratio is close to unity and is similar to what is found in local star-forming systems. Based on low-redshift observations, we would have expected a higher ratio for low-metallicity galaxies or a lower ratio in AGN-dominated systems.
\item After deblending the HNC(5-4) and CN(4-3) lines, we found an average I$_{\rm HNC(5-4)}$/I$_{\rm HCN(5-4)}$ flux ratio of $1.03_{-0.39}^{+0.59}$, placing our objects at the border between the PDR-dominated regime corresponding to a ratio below unity and the high-ratio regime caused by XDR and/or mid-IR pumping of HNC.
\item We inferred the HCN(5-4)/[CI](1-0) ratio for our sources and found it to be higher
than what is measured in LIRGs and ULIRGs in the local Universe. To the extent that this ratio is a proxy of the dense-gas
fraction, this finding would suggest that our SPT sources have a larger dense-gas fraction than local (U)LIRGs.
In the diagram showing the dense-gas fraction versus the star formation efficiency relative to the bulk
gas mass, our sources are located on the same linear relation as the local galaxies. 
\item In SPT0125-47, we compared the profiles of our 5.5-$\sigma$ detections of HCN(5-4) and HCO$^{+}$(5-4), our 36-$\sigma$ detection of [CI](1-0), and the CO(4-3) line from the original redshift-search observation. All these lines have the same asymmetric profile. This suggests that the low- and high-density molecular gas is distributed in the same way across this galaxy. This similarity of the line profiles also suggests the absence of significant differential lensing.
\item Concerning the CO isotopologues, our sample has a mean I$_{\rm ^{12}CO(4-3)}$/I$_{\rm ^{13}CO(4-3)}$ flux ratio of 26.1$_{-3.5}^{+4.5}$, which is 30\% lower than the previously observed high-z galaxies but 50\% higher than typical local LIRGs. Even if the interpretation of this single line ratio is ambiguous, it tends to indicate that the ISM of our objects has already been enriched by intermediate-mass stars.
\end{itemize}
Even if the interpretation of these lines is non-trivial, our study allows
us to draw a first preliminary portrait of the dense ISM in the SPT SMG sources.
Previous works had already established that they had large gas reservoirs
\citep{Aravena2016}, but we can now confirm that they have also a high DGF, as traced
by HCN(5-4)/[CI](1-0). A high DGF naturally leads to a higher star formation efficiency
measured relative to the bulk gas mass, that is, a high L$_{\rm IR}$/L'$_{\rm CO(1-0)}$ or, as in our case, a
high L$_{\rm IR}$/ L'$_{\rm [CI](1-0)}$. The combination of large reservoirs and
high SFEs might be sufficient to explain the prodigious SFRs observed for the SPT sources. The ratios between dense-gas
tracers are compatible with PDR-dominated objects and no evidence of AGN
activity has been found so far. 

Combining our sample with the previously observed high-redshift galaxies
dominated by star formation, we currently have dense-gas data for less than ten
systems. Our analysis revealed some small tensions with other studies. However,
their statistical significance is often below 2-$\sigma$. We need larger samples
and higher S/N ratios to identify whether these differences are real and if they are linked
to the physical properties of the galaxies. Our study shows that dense-gas
tracers can be detected in $\sim$1\,h with ALMA in highly-magnified objects and
we can realistically hope to build a sample containing 20-30 objects in the
coming years. Unfortunately, HCN is a bit fainter than expected from the
observations at low redshift, but this bad news is totally compensated for by the
fact that HCO$^{+}$ and HNC are at least as bright and can be detected by the
same observations.

\begin{acknowledgements}
MB would like to thank Zhi-Yu Zhang, Chentao Yang, and Alain Omont for providing their data points and/or the insightful discussions that we had about dense-gas tracers. This paper makes use of the following ALMA data: ADS/JAO.ALMA\#2016.1.00065.S, ADS/JAO.ALMA\#2011.0.00957.S. ALMA is a partnership of ESO (representing its member states), NSF (USA), and NINS (Japan), together with NRC (Canada), MOST and ASIAA (Taiwan), and KASI (Republic of Korea), in cooperation with the Republic of Chile. The Joint ALMA Observatory is operated by ESO, AUI/NRAO, and NAOJ. MA acknowledges partial support from FONDECYT through grant 1140099.
\end{acknowledgements}

\bibliographystyle{aa}

\bibliography{biblio}

\begin{appendix}

\section{High-S/N detection of [CI](1-0) in SPT0125-47}

\label{sect:ci}

\begin{figure}
\centering
\includegraphics[width = 9cm]{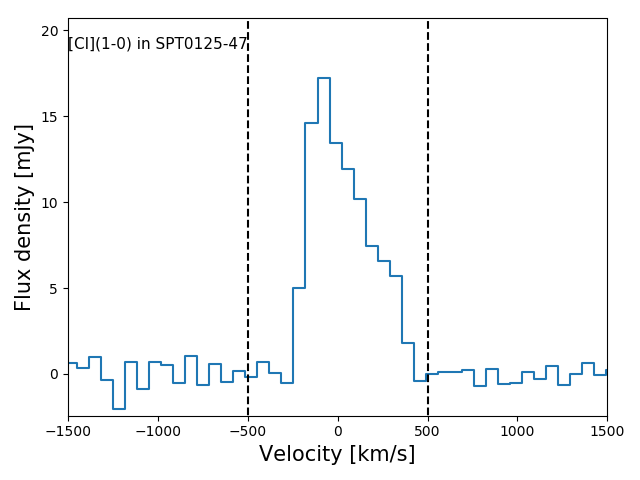}
\caption{\label{fig:CI} Detection of [CI](1-0) in SPT0125-47. The continuum has been subtracted from the spectrum. The vertical dashed lines indicate the window in which the spectrum was integrated to derive the line flux presented in Table\,\ref{table:ancillary_lines}.}
\end{figure}

We detected the [CI](1-0) line of SPT0125-47 for first time. This relatively bright line was not covered by the original band-3 spectral scan used to determine the redshift of the source, but is located in the other side band of our new, deep ALMA band-4 observations of dense-gas lines. We produced a spectrum using the  same method as in Sects.\,\ref{sect:data_red} and \ref{sect:line_ext}. The line is detected at a very high S/N and exhibits an asymmetric profile (see Fig.\,\ref{fig:CI}), making a Gaussian model not appropriate. We measured the flux in a window set manually and covering the full line emission. We derived the uncertainties using the standard deviation outside the frequency window, where the line is located. The line is detected with an S/N of 36. Table\,\ref{table:ancillary_lines} contains the [CI](1-0) flux of SPT0125-47.

\begin{table*}
\caption{\label{tab:mom0_comp} Comparison between our measurements preformed using the method described in Sect.\,\ref{sect:line_ext} and the classical approach based on the moment-zero map (Appendix\,\ref{sect:fit_vs_mom0}).}
\centering
\begin{tabular}{llrrrr}
\hline
\hline
Source & Line & \multicolumn{2}{c}{Flux density (Jy\,km/s)}  & \multicolumn{2}{c}{S/N}\\
& & Fit + bootstrap & moment-zero & Fit + bootstrap\tablefootmark{a}  & moment-zero\\
\hline
SPT0125-47 & $^{13}$CO(4-3) & 1.15$\pm$0.22 & 0.90$\pm$0.15 & 7.2 & 6.0\\
 & HCN(5-4) & 0.96$\pm$0.23 & 0.91$\pm$0.15 & 5.3 & 6.0\\
 & HCO$^{+}$(5-4) & 1.01$\pm$0.20 & 0.91$\pm$0.14 & 5.6 & 6.5\\
 & [CI](1-0) & 6.36$\pm$0.27 & 6.21$\pm$0.13 & 27 & 41\\
\hline
SPT0551-50 & $^{13}$CO(4-3) & 0.32$\pm$0.15 & 0.37$\pm$0.07 & 3.4 & 5.3\\
 & HCN(5-4) & 0.22$\pm$0.11 & 0.25$\pm$0.08 & 3.0 & 3.2\\
 & HCO$^{+}$(5-4) & 0.17$\pm$ 0.07 & 0.24$\pm$0.07 & 3.1 & 3.4\\
 & HNC+CN & 0.25$\pm$ 0.09 & 0.26$\pm$0.08 & 2.6 & 3.3\\
\hline
\end{tabular}
\tablefoot{\tablefoottext{a}{The S/N of our fitting method is determined by computing S$_{\rm peak}$ / $\sigma_{S_{\rm peak}}$.}}
\end{table*}

\section{Checking the consistency between our line-extraction method and the classical moment-zero map approach}

\label{sect:fit_vs_mom0}

Our extraction method presented in Sect.\,\ref{sect:line_ext} relies on the quality of the Gaussian assumption for the line profile. The [CI](1-0) profile shown in Appendix\,\ref{sect:ci} has a very high S/N and is clearly non-Gaussian. Usually, the best way to deal with non-Gaussian lines is to produce a moment-zero map by summing all the channels, which contains flux from the lines. However, this method works only if the continuum has been previously subtracted in the uv plane (\textit{uvcontsub} in CASA).

Unfortunately, this subtraction is difficult in the case of our data, since we can only use  a couple of clean continuum channels between HCO$^{+}$(5-4) and HNC(5-4) in the side band of the dense-gas tracers. This is caused by the presence of several lines in this side band, which can be particularly broad, especially in SPT0551-50. The line-versus-continuum flux density ratio is of the order of unity and the slope of the continuum versus frequency inside a spectral window is non-negligible. This makes this task even more difficult. We performed various tests and realized that the continuum subtraction was very sensitive to our parameter choices, creating important systematics. For instance, with SPT0551-50, the recovered flux in the moment-zero map varies by more than 1-$\sigma$ if we remove a flat background measured between HCO$^{+}$(5-4) and HNC(5-4) or if we use a first-order baseline using in addition the other side band to constrain the slope. This is even worse for the three sources at low S/N (SPT0103-45, SPT0125-50, SPT0300-46) for which the flux of the line can vary by a factor of two depending on how the continuum was subtracted.

These problems motivated our choice to use our own method introduced in Sect.\,\ref{sect:line_ext} to derive line fluxes. However, we performed a comparison between our approach and the moment-zero measurements using a first-order continuum subtraction. We restrict this comparison to SPT0125-47 and SPT0551-50 for which we can clearly identify the channels contaminated by the lines and thus reliably subtract the continuum. The results are summarized in Table\,\ref{tab:mom0_comp}. The values derived by the two methods are compatible at 1\,$\sigma$ for all the lines. The uncertainties derived using our own method are slightly higher than the ones obtained with the moment maps. This is expected, since it takes into account the impact of the uncertainties on the baseline level and on the line width, while the moment-zero method assumes implicitly that the continuum is subtracted perfectly and uses a fixed velocity window to integrate the line flux. While they are slightly higher, the uncertainties provided by our fitting method are thus probably more realistic. Concerning the S/N ratios, the estimates based on the moment-zero maps are close to the ones derived using the ratio between the peak line flux density determined with our method and the uncertainty on it (S$_{\rm peak}$ / $\sigma_{S_{\rm peak}}$).

In SPT0125-47, the S/N and the line flux of [CI](1-0) based on the method presented in Appendix\,\ref{sect:ci} is between the moment-zero and the fitting method and we will keep this intermediate value for analysis. Indeed, since the [CI] flux is used to compute ratios with much more uncertain quantities, these small differences do not have any significant impact on our results. Our Gaussian fit and the moment-zero method agree at the 3\% level. This confirms that the Gaussian approximation is a reasonable assumption for lines detected at $\sim$5\,$\sigma$. In this case, the systematic effect induced by the non-Gaussian profile of [CI] is only 0.15\,$\sigma$.

\begin{table}
\centering
\caption{\label{tab:simu} Results of our simulation validating our method to derive mean luminosities and ratios (see Appendix\,\ref{sect:average_simu}). For line fluxes, the true sample mean value $\langle X_{\rm true} \rangle$ is the mean of the 500 injected fluxes of our simulated sample. As justified in Sect.\,\ref{sect:unbiased_ratio}, the mean line flux ratios are derived by dividing the mean flux of a given line by the mean flux of another line. The fluxes measured using our line extraction tool are used to derive $\langle X_{\rm mes} \rangle$. $\frac{\langle X_{\rm mes} \rangle}{\langle X_{\rm true} \rangle}$ must be close to unity if the method is not biased. $\frac{\sigma_{\rm full} (\langle X_{\rm mes} \rangle)}{\langle X_{\rm true} \rangle}$ is the uncertainty on this ratio caused by the instrumental noise. Finally, by drawing 1000 subsamples of five sources from our 500 objects without withdrawal, we estimated the mean uncertainty on the measured-versus-true ratio for a subsample of five objects $\frac{\sigma_{5} (\langle X_{\rm mes} \rangle)}{\langle X_{\rm true} \rangle}$ as is the case for our real observations.}
\begin{tabular}{lrrrrr}
\hline
\hline
Observable & $\langle X_{\rm true} \rangle$ & $\langle X_{\rm mes} \rangle$ & $\frac{\langle X_{\rm mes} \rangle}{\langle X_{\rm true} \rangle}$ & $\frac{\sigma_{\rm full} (\langle X_{\rm mes} \rangle)}{\langle X_{\rm true} \rangle}$ & $\frac{\sigma_{5} (\langle X_{\rm mes} \rangle)}{\langle X_{\rm true} \rangle}$ \\
\hline
\multicolumn{5}{c}{Line fluxes in Jy\,km/s} \\
\hline
I$_{\rm ^{13}CO(4-3)}$ & 0.148 & 0.157 & 1.063 & 0.016 & 0.183 \\
I$_{\rm HCN(5-4)}$ &       0.142 & 0.146 & 1.030 & 0.015 & 0.177 \\
I$_{\rm HCO^{+}(5-4)}$ & 0.149 & 0.155 & 1.038 & 0.016 & 0.158 \\
I$_{\rm HNC(5-4)}$ &       0.149 &  0.152 & 1.017 & 0.017 & 0.172 \\
I$_{\rm CN(4-3)}$ &          0.093 &  0.091&  0.974 & 0.026 & 0.263 \\
\hline
\multicolumn{5}{c}{Line flux ratios} \\
\hline
$\frac{I_{\rm HCO^{+}(5-4)}}{I_{\rm HCN(5-4)}}$ & 1.050 & 1.059 & 1.008 & 0.020 & 0.326 \\
$\frac{I_{\rm HNC(5-4)}}{I_{\rm CN(4-3)}}$ &          1.596 & 1.672 & 1.047 & 0.116 & 1.559 \\
$\frac{I_{\rm HNC(5-4)}}{I_{\rm HCN(5-4)}}$ &        1.055 & 1.042 & 0.988 & 0.031 & 0.452 \\
\hline
\end{tabular}
\end{table}

\section{Validation of our method deriving mean luminosities and ratios}

\label{sect:average_simu}

In Sects.\,\ref{sect:unbiased_ratio} and \ref{sect:deblend_ratio}, we presented a statistical approach to derive mean luminosities and line ratios without being biased towards detections. In this appendix, we present the simulation used to validate our method.

We generated 500 simulated spectra at z$\sim$3.2 of two spectral windows covering $^{13}$CO(4-3), HCN(5-4), HCO$^{+}$(5-4), HNC(5-4), and CN(4-3) (104.6--108.35\,GHz). This redshift was chosen because it allows us to cover all the line at the same time and it is close to the mean redshift of the real sample. With these 500 spectra in hand, we reduced the statistical uncertainties on the derived mean quantities by a factor of ten compared to our real sample. This allows us to check if small systematic biases are affecting our approach.

For each source, we drew the fluxes of the $^{13}$CO(4-3), HCN(5-4), HCO$^{+}$(5-4), and HNC(5-4) lines from an uniform distribution between 0\,Jy\,km/s and 0.3\,Jy\,km/s, which is the typical range in our real sample if we leave out the exceptionally bright SPT0125-47. We used a maximum flux of 0.187\,Jy\,km/s for CN(4-3) to be consistent with the mean observed flux ratio with HNC(5-4) (1.6, see Sect.\,\ref{sect:deblend_ratio}). For all the lines of a given source, we used the same line width drawn from a uniform distribution centered on 370\,km/s and with a half width of 130\,km/s as measured by \citet{Aravena2016} for low-J CO lines in the SPT SMG sample. Finally, we added a random Gaussian noise of 0.2\,mJy per channel of 31.25\,GHz (resolution of SPT0125-47 after rebinning). We  extracted the lines in these simulated spectra using the method presented in Sect.\,\ref{sect:line_ext}. For HNC(5-4) and CN(4-3), following Sect.\,\ref{sect:deblend_ratio}, we performed a second run assuming a positivity prior and no velocity offset to obtain a better deblending. The median S/N of the $^{13}$CO(4-3), HCN(5-4), and HCO$^{+}$(5-4) lines is 2.9,  2.8, and 3.1, respectively. Roughly half of our simulated sample is thus not even tentatively detected.

In Table\,\ref{tab:simu}, we compare the true fluxes injected into our simulation with the measured ones. The maximal difference between these two quantities is 6.3\,\%. This demonstrates that our method is reasonably accurate to derive the mean flux of a sample even if a large fraction of the objects is not detected. This difference of 6.3\,\% obtained for $^{13}$CO(4-3) is significant at 4\,$\sigma$, when we consider the uncertainty corresponding to the instrumental noise in our simulated spectra. There are thus residual systematic effects, but they remain small compared to the typical uncertainties on our measurements based on four or five sources ($\gtrsim$20\%). In addition, we found that the measured fluxes of $>3\,\sigma$ (tentative) detections are overestimated on average by 28\,\%, 41\,\%, and 13\,\% for $^{13}$CO(4-3), HCN(5-4), and HCO$^{+}$(5-4), respectively. Our statistical approach is thus clearly better than computing naively the mean flux of the detected sources.

Finally, we compared the flux ratio derived using our statistical method and the true ratio (Table\,\ref{tab:simu}). We  found no significant deviation. The small systematic effects affecting mean line fluxes could compensate for each other. The ratios including blended lines are also well recovered. This is a good demonstration of the accuracy of our approach including for blended lines (Sect.\,\ref{sect:deblend_ratio}).

\begin{figure}
\centering
\includegraphics[width=9cm]{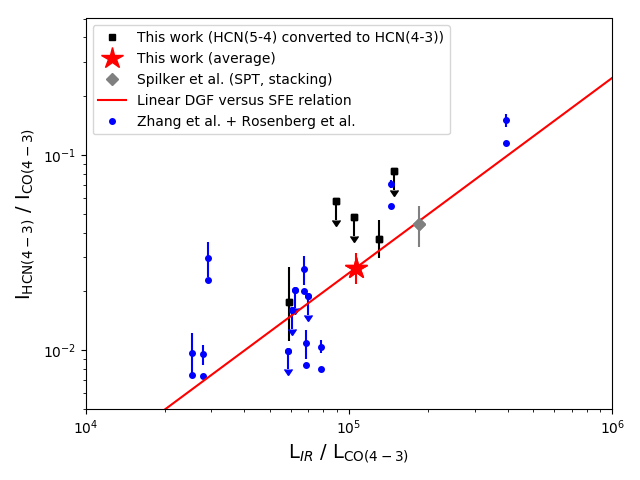}
\caption{\label{fig:sfe_co} Ratio between HCN(4-3) and CO(4-3), tracing the dense-gas fraction, as a function of the ratio between the infrared and the CO(4-3) luminosity, tracing the star formation efficiency. The red solid line is a linear relation between the dense-gas fraction and the star formation efficiency (DGF $\propto$ SFE), which normalization has been set to match the mean value of our sample.}
\end{figure}

\begin{figure}
\includegraphics[width=9cm]{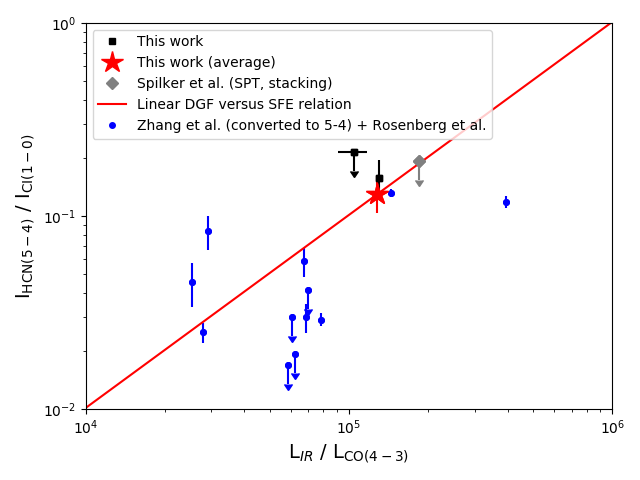}
\includegraphics[width=9cm]{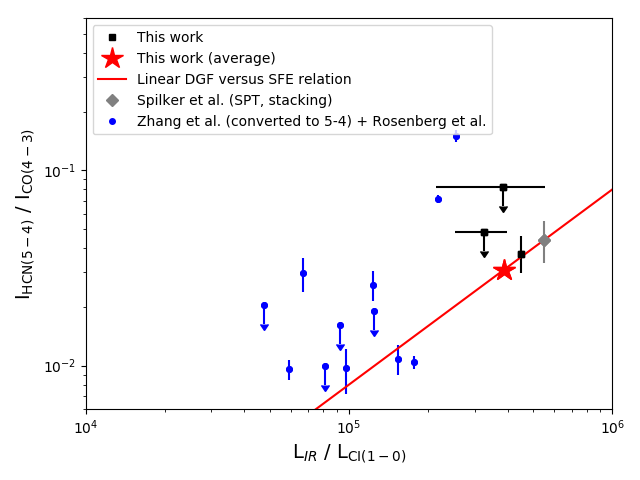}
\caption{\label{fig:sanity_dgf_sfe} Upper panel: Dense-gas fraction as a function of star formation efficiency. The data are the same as in Fig.\,\ref{fig:sfe}, except that we used [CI](1-0) instead of CO(4-3) on the y-axis. Lower panel: Same thing as upper panel, except that we switched CO(4-3) and [CI](1-0) on the x and y axes.}
\end{figure}

\section{Confirming the correlation between dense-gas fraction and star formation efficiency}

\label{sect:sanity}

In Fig.\,\ref{fig:sfe_co}, we present a version of Fig.\,\ref{fig:sfe} using CO(4-3) instead of [CI](1-0). Similarly to the original figure, our results are compatible with a linear relation between DGF and SFE. CO(4-3) is not a very good tracer of the total gas mass, since it is biased towards denser environments compared to CO(1-0) and it is affected by the variations of the SLED. However, it is reassuring to find a similar result with this independent tracer.

In Sect.\,\ref{sect:dgf_sfe} and Fig.\,\ref{fig:sfe}, we discussed the link between the star formation efficiency and the dense-gas fraction using a diagram of the form A/C versus B/C. In this case, C is a quantity tracing the total molecular gas reservoir of the galaxy. Such diagrams can produce artificial correlations, if the uncertainties on C are of the order of magnitude of A/C or B/C. This is not expected in our diagram, where we probe more than one order of magnitude on both axes, while the measurements uncertainties on the [CI](1-0) and CO(4-3) fluxes are lower than 30\,\%. We tested whether the trend of a DGF increasing with higher SFE remains when we use two different tracers of the total cold gas on each axis (Fig.\,\ref{fig:sanity_dgf_sfe}). There is a slightly larger scatter than in the previous plots, but the trend does not disappear. This larger scatter could come from the addition of the systematic effects affecting each tracer in a different way.

\section{Biased $^{13}$CO/$^{12}$CO ratios in samples of serendipitous $^{13}$CO detections}

\label{sect:bias_serendipitous}

In Sect.\,\ref{sect:isotopes}, we found that serendipitous $^{13}$CO detections in the SPT SMG redshift-search program of \citet{Weiss2013} have much lower $^{12}$CO/$^{13}$CO ratios than both our detections and the stacking of \citet{Spilker2014}. To understand this effect, we performed a simplified simulation.

We started from the measured fluxes of the 29 $^{12}$CO lines detected in band 3 in the current SPT SMG sample. These lines are so bright that they are always detected at high S/N and we can thus neglect their flux measurement uncertainties. We then generated a "true" $^{13}$CO flux by dividing the $^{12}$CO by a factor of 26 (the mean line flux ratio found for our sample) and applying a log-normal scatter of 0.3\,dex, which is approximately what is seen in Fig.\,\ref{fig:isotope_comp}. We also took into account the impact of the instrumental noise on the measurements of the $^{13}$CO fluxes. The noise can vary from one source to  another. For simplicity, we assumed the same Gaussian noise $\sigma_{\rm noise}$ of 0.365 Jy\,km/s for all the sources, which is the mean of the noise of the two \citet{Weiss2013} detections. We consider as detected only the objects with a "measured" flux larger than 3\,$\sigma_{\rm noise}$. Since the number of these detections can vary significantly from one realization of the intrinsic scatter and the instrumental noise to the other, we repeated this operation 100\,000 times to minimize the statistical fluctuations. We found on average 3.3 detections with a standard deviation of 1.6. This is compatible with the two detections in the real sample of \citet{Weiss2013}.

By construction, the mean $^{12}$CO/$^{13}$CO ratio of our complete simulated catalog is 26. In contrast, the mean "true" ratio of the detections is 11.3 with standard deviation of 7.6. The subsample detected in $^{13}$CO is thus strongly biased towards low ratios. The mean ratio is even smaller when we compute it using the "measured" $^{13}$CO flux: 8.1 with scatter of 4.5. The instrumental noise thus increases the bias, since the $^{13}$CO detections are affected by a severe flux boosting. We found a mean ratio of 1.5 between the "measured" and the "true" flux. These values of the measured ratio are very close to the two serendipitous detections of \citet{Weiss2013}.

This simulation demonstrates that the rare serendipitous detections extracted from large samples can be highly biased and should be used with caution. It also shows why it is so important to derive average properties of samples considering also the non-detections to interpret results from a sample (see Sect.\,\ref{sect:unbiased_ratio}).

\end{appendix}

\end{document}